\documentclass[%
 aip,
citeautoscript,
 amsmath,amssymb,
 preprint,%
]{revtex4-2}

\usepackage{graphicx}
\usepackage{lipsum}
\usepackage{ccaption}
\usepackage{dcolumn}
\usepackage{bm}

\usepackage[utf8]{inputenc}
\usepackage[T1]{fontenc}
\usepackage{mathptmx}
\usepackage{etoolbox}
\usepackage{rotating}

\usepackage[version=4]{mhchem}
\usepackage{siunitx}
\DeclareSIUnit\au{a.u.}
\usepackage{comment}
\usepackage{import}
\usepackage{multirow}
\usepackage{xcolor}
\usepackage{braket}

\allowdisplaybreaks

\newcommand{\kJmol}[0]{kJ~mol$^{-1}$}
\newcommand{\kcalmol}[0]{kcal~mol$^{-1}$}

\makeatletter
\def\@email#1#2{%
 \endgroup
 \patchcmd{\titleblock@produce}
  {\frontmatter@RRAPformat}
  {\frontmatter@RRAPformat{\produce@RRAP{*#1\href{mailto:#2}{#2}}}\frontmatter@RRAPformat}
  {}{}
}%
\makeatother
\begin{document}

\title[]{Prospects for rank-reduced CCSD(T) in the context of high-accuracy thermochemistry}
\author{Tingting Zhao}
 \affiliation{Department of Chemistry, Southern Methodist University, Dallas, TX}
\author{James H. Thorpe}%
 \affiliation{Department of Chemistry, Southern Methodist University, Dallas, TX}

\author{Devin A. Matthews*}
 \affiliation{Department of Chemistry, Southern Methodist University, Dallas, TX}
 \email{damatthews@smu.edu}


\begin{abstract}
Obtaining sub-chemical accuracy (1 \kJmol) for reaction energies of medium-sized gas-phase molecules is a longstanding challenge in the field of thermochemical modeling. The perturbative triples correction to CCSD, CCSD(T), constitutes an important component of all high-accuracy composite model chemistries that obtain this accuracy, but can be a roadblock in the calculation of medium to large systems due to its $\mathcal{O}(N^7)$ scaling, particularly in HEAT-like model chemistries that eschew separation of core and valance correlation. This study extends the work of Lesiuk [\emph{J. Chem. Phys.} \textbf{156}, 064103 (2022)] with new approximate methods and assesses the accuracy of five different approximations of (T) in the context of a subset of molecules selected from the W4-17 dataset. It is demonstrated that all of these approximate methods can achieve sub-0.1 \kJmol accuracy with respect to canonical, density-fitted (T) contributions with a modest number of projectors. The approximation labeled $\tilde{Z}T$  appears to offer the best trade-off between cost and accuracy and shows significant promise in an order-of-magnitude reduction in the computational cost of the CCSD(T) component of high-accuracy model chemistries.
\end{abstract}

\maketitle

\section{\label{sec:Introduction}Introduction}


Wavefunction-based composite model chemistries are the leading methods of choice for obtaining reaction energies and molecular enthalpies of formation. These techniques assume that the solution of the ``true'' Schr\"{o}dinger equation for a particular molecule can be represented as a sum over (approximately) separable contributions to the energy. These typically include, in order of importance, the non-relativistic electronic energy, the (ro)vibrational zero point energy, corrections for scalar relativistic and spin-orbit effects, and the so-called diagonal Born-Oppenheimer correction. The focus of this work, the non-relativistic electronic energy (NREE), is generally the largest contributor for a particular observable\cite{Karton2006, Feller2008, Harding2008}. This term is usually obtained via a further series of additive coupled-cluster calculations, where the effects of higher-order cluster operators are treated with monotonically decreasing basis sets to offset the computational cost. There are few families of such high-accuracy model chemistries that fall into one of two camps: ``fixed'' recipes such as W$n$\cite{Boese-JCP-2004, Karton2006, Sylvetsky2016}, HEAT\cite{Tajti2004, Harding2008, Bomble2006, Thorpe2019, Thorpe-JCP-2021}, and ANL$n$\cite{Klippenstein2017}, which prescribe a fixed protocol and obtain Type-A uncertainties\cite{Ruscic-IJQC-2014}, and ``free'' recipes such as FPA\cite{East1993, Csaszar-JCP-1998, Kenny-JCP-2003, Schuurman-JCP-2004} and FPD\cite{Feller2008, Feller2012, Peterson2012, Dixon2012, Feller-TCA-2013} which provide guidelines for constructing model chemistry for a molecule of interest and obtain Type-B uncertainties\cite{Ruscic-IJQC-2014}. 

While the particulars of these calculations differ from family to family, it is of particular note that every single one of them employs CCSD(T)\cite{T_Raghavachari1989} as an additivity point in their recipes, often as a component of a $T_3$ correction to the $T_1 + T_2$ correlation obtained by CCSD\cite{Purvis-JCP-1982} (the exception to this statement is the HEAT family of methods, which avoid separation of CCSD and CCSD(T) correlation). CCSD(T) provides a felicitous compromise between accuracy and computational cost and goes a long way towards recovering post-CCSD correlation with basis sets large enough to provide the accuracy required of these model chemistries. This last point is important: CCSD(T) scales as $\mathcal{O}(N^7)$, where $N$ is the number of molecular orbitals, and can be one of the most expensive parts of these model chemistries for larger systems when large basis sets are required, particularly in HEAT-like methods which do not separate the basis-set dependence of CCSD from its perturbative (T) corrections.


Thanks to advances in hardware and software of supercomputers, CCSD(T) can now be applied to molecules with over twenty carbons using medium-size basis sets like cc-pVTZ.\cite{Karton2013, Karton2015, Manna2015} Further, the development of localized techniques has empowered the calculation of (T)-like corrections for systems containing hundreds of atoms, and have been shown to approach \kcalmol accuracy.\cite{Riplinger2013_jan, Riplinger2013_oct, Liakos-JCTC-2015, Ma2018, Nagy2019}  However,  sub-chemical accuracy usually requires (T) contributions calculated with basis sets of at least QZ size (ideally 5Z or higher)\cite{Karton2007, Feller2011, Feller2013}, and such accuracy places significant strain on localized methods for calculating this quantity. The focus of this paper is to explore how the scaling of this correction can be reduced in the context of high-accuracy computational thermochemistry.  

There are four types of terms that can be usefully approximated in the calculation of the perturbative corrections in CCSD(T): the two-electron repulsion integrals (ERIs), the two- and three-electron cluster amplitudes ($T_2$ and $T_3$), and the three-electron orbital eigenvalue denominators ($D_3$). Each of these has been studied individually in the literature, and we provide a brief overview of the relevant work here. Density fitting via the Coulomb metric has long been used to approximate the two-electron ERIs\cite{boys1959, Vahtras1993, Rendell1994}. Of the methods discussed here, this is probably the best understood and most well characterized, and for most applications, including this one, results in little-to-no appreciable error in predicted properties. Approximation of the orbital eigenvalue denominator in CCSD(T) via an (inverse) Laplace transform\cite{almlof1991,Haser1992} quadrature\cite{braessApproximationExponentialSums2005} was first studied by Constans et al.\cite{Constans2000}, who demonstrated that the canonical $\mathcal{O}(N^{7})$ scaling could be reduced to $\mathcal{O}(N^{6})$, although the leading term still scales unfavorably with the fifth power of the number of virtual orbitals. The Laplace transform can be made essentially exact with a sufficient number of quadrature points, though this results in an increase in the numerical prefactor of the computational cost. In the approximation of $T_2$, Kinoshita, Hino, and Bartlett demonstrated that the CCD equations could be represented in an SVD subspace, effectively ``compressing'' the two-electron amplitudes of the CC equations.\cite{Kinoshita2003} Parrish et al. later presented a ``rank-reduced'' coupled cluster (RR-CC) method for compression of $T_2$, where the compressed doubles amplitudes are solved directly via a Lagrangian formulation of the CCSD equations.\cite{Parrish2019} These works demonstrated that the $T_2$ amplitudes were of significantly lower rank than their formal sizes; in fact, it appears that number of significant singular values of $T_2$ scales only linearly with system size. Further, as the compression vectors used in these rank-reduced methods form an orthogonal basis, orbital rotation can be used to simplify expressions for the orbital energy denominators.\cite{Hino2004,Parrish2019} Similar attempts to approximate the $T_3$ amplitudes were explored by Hino, Kinoshita, and Bartlett in the context of CCSDT-1\cite{Urban-JCP-1985, Hino2004}, but the computational cost of decomposing $T_3$ remained prohibitively expensive until developments utilizing Golub-Kahan bidiagonalization and/or higher order orthogonal iteration (HOOI) reduced the cost of decomposing $T_3$-like quantities to $\mathcal{O}(N^6)$.\cite{Lesiuk2019, Lesiukacs2019, Lesiuk2022} Of particular relevance is the recent work of Lesiuk, who used HOOI to reduce the cost of the entire CCSD(T) correction to a leading-order scaling the same as CCSD ($\mathcal{O}(o^2 v^4)$, where $o$ and $v$ are the number of occupied and virtual orbitals, respectively).\cite{Lesiuk2022} Importantly, that manuscript reported less than 1\% errors in the compressed CCSD(T) isomerization energies, which makes it a prime candidate for application in model chemistries aiming for sub-chemical accuracy. 

In this work we explore several combinations of the aforementioned techniques, including several novel factorization strategies, and to attempt to gain some insight into the application of these reduced scaling techniques in high-accuracy thermochemistry. It should be noted that other decompositions of the electron repulsion integrals and CC amplitudes are possible---of particular note are Tensor Hypercontraction methods\cite{Hohenstein-JCP-2012, Matthews-JCP-2021, Jiang2023}---but as the focus of this work is on high-accuracy thermochemistry we have elected to focus only on the set of techniques discussed above. 

\section{Theory}

Different methods of approximating the (T) correction and their scaling will be discussed in detail in this section. Throughout this work, the following notational conventions are used:
\begin{itemize}
\setlength{\baselineskip}{0pt}
\item The letters $ijkl$ represent one of the $o$ occupied orbitals in the molecular orbital (MO) basis;
\item The letters $abcd$ represent one of the $v$ virtual orbitals in the MO basis;
\item The letters $pqrs$ represent one of the $N = o+v$ general orbitals in the MO basis, which may be fewer than the number of basis functions $N_{bas}$ if orbitals are frozen;
\item The letter $g$ represents one of the $N_g$ Laplace transform quadrature points;
\item The letters $VW$ represent one of the $N_{T_2}$ rank-reduced $T_2$ projectors;
\item The letters $XYZABC$ represent one of the $N_{T_3}$ Tucker-3 compressed $T_3$  projectors;
\item The letters $JKL$ represent one of the $N_{DF}$ density-fitting auxiliary basis functions.
\end{itemize}

\subsection{Perturbative triples correction to CCSD}

As a starting point, we take the coupled cluster singles and doubles (CCSD) energy $E_{CC}$ and amplitudes $t_n$,
\begin{align}
E_{CC} &= \langle 0|\bar{H}|0\rangle \nonumber \\
&= \langle 0|e^{-\hat{T}} \hat{H}_N e^{\hat{T}} |0\rangle \nonumber \\
&= \langle 0|(\hat{H}_N e^{\hat{T}})_C |0\rangle \\
\hat{H}_N &= \hat{F}_N + \hat{V}_N \nonumber \\
&= \sum_{pq} f^p_q \{a^\dagger_p a_q\} + \frac{1}{2} \sum_{pqrs} v^{pq}_{rs} \{a^\dagger_p a^\dagger_q a_s a_r\} \\
\hat T &= \hat{T}_1 + \hat{T}_2 \nonumber \\
&= \sum_{ai} t^a_i a^\dagger_a a_i + \frac{1}{4} \sum_{abij} t^{ab}_{ij} a^\dagger_a a^\dagger_b a_j a_i
\end{align}
where we use the non-antisymmetrized two-electron integrals $v^{pq}_{rs} = \langle p q| rs\rangle = ( p r| qs )$. From this point, we will assume the use of a restricted Hartree-Fock (RHF) reference and refer only to spatial orbitals. In CCSD(T), $E_{CC}$ is supplemented by two terms arising from triple excitations: the so-called fourth-order correction, $E_{T}^{[4]}$, and the so-called fifth-order correction, $E_{ST}^{[5]}$,\cite{Pople1987}
\begin{align}
E_{(T)}&=E_{T}^{[4]}+E_{ST}^{[5]}  \\
E_{T}^{[4]}&=\frac{1}{3}\sum_{abcijk}z_{ijk}^{abc}S_{ijk}^{abc}t_{ijk}^{abc}  \\
E_{ST}^{[5]}&=\frac{1}{6}\sum_{abcijk}\left( P^{abc}_{ijk}v_{ij}^{ab}t_{k}^{c}\right) S_{ijk}^{abc}t_{ijk}^{abc} \\
z_{ijk}^{abc}&=P_{ijk}^{abc}\left( \sum_{d}v_{kd}^{cb}t_{ij}^{ad}-\sum_{l}v_{kj}^{cm}t_{im}^{ab} \right) \\
t_{ijk}^{abc}&=z_{ijk}^{abc}D_{ijk}^{abc}
\end{align}
$D_{ijk}^{abc}$ are the orbital eigenvalue denominators and $P_{ijk}^{abc}$ and $S_{ijk}^{abc}$ are permutation operators,
\begin{align}
D_{ijk}^{abc}&=\frac{1}{({\epsilon}_{i}+{\epsilon}_{j}+{\epsilon}_{k}-{\epsilon}_{a}-{\epsilon}_{b}-{\epsilon}_{c})} \\
P_{ijk}^{abc}&=\left(_{ijk}^{abc}\right)+\left(_{ikj}^{acb}\right)+\left(_{jik}^{bac}\right)+\left(_{kij}^{cab}\right)+\left(_{jki}^{bca}\right)+\left(_{kji}^{cba}\right) \\
S_{ijk}^{abc}&=4\left(_{ijk}^{abc}\right)-2\left(_{ijk}^{acb}\right)-2\left(_{ijk}^{bac}\right)-2\left(_{ijk}^{cba}\right) +\left(_{ijk}^{bca}\right)+\left(_{ijk}^{cab}\right)
\end{align}
The unfavorable $\mathcal{O}(N^{7})$ scaling comes from the construction of $z_{ijk}^{abc}$ from the two-electron integrals and amplitudes, and approximating this term is particularly important in reducing the cost of CCSD(T). We will use the $E_T^{[4]}$ term (diagrammatically represented in Fig.~\ref{fig:Methods_proposed}A) to demonstrate the various approximate methods explored here. Scaling reduction of $E_{ST}^{[5]}$ is achieved using the same principles as $E_T^{[4]}$, and $E_{ST}^{[5]}$ is included in the numerical results presented below.


\begin{figure}
    \centering
    \includegraphics[width=0.45\textwidth]{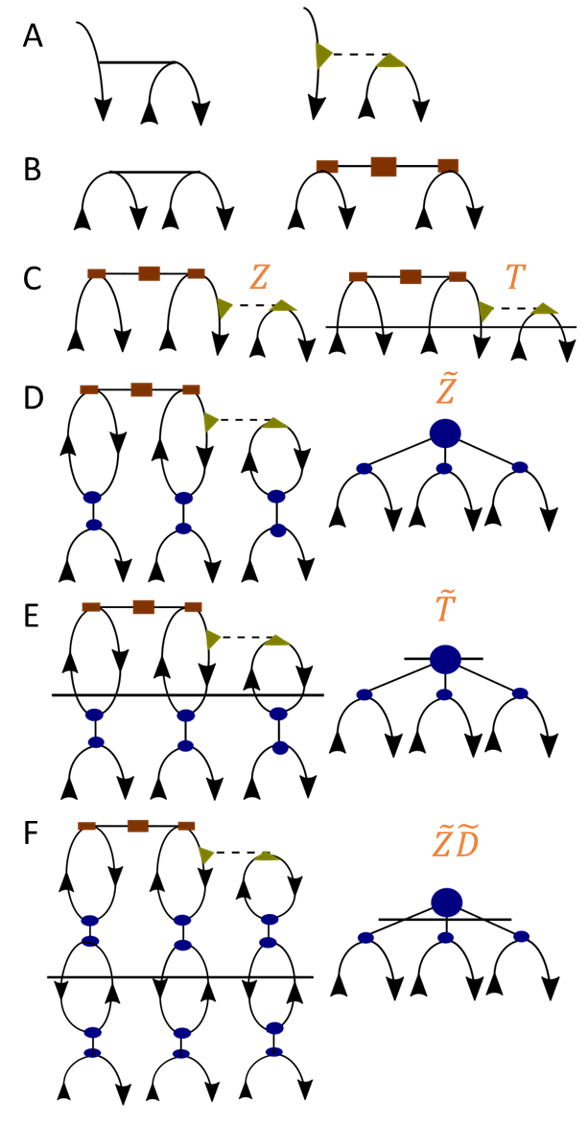}
    \caption{Graphical representations of the various decompositions employed. See text for details. \textbf{(A)} Left: ERIs ($v_{kd}^{cb}$ or $v_{kj}^{cm}$), right: DF-approximated ERIs; \textbf{(B)} Left: $T_2$ amplitudes, right: rank-reduced $T_2$ amplitudes; \textbf{(C)} Left: $z_{ijk}^{abc}$ approximated with DF and RR ($\bar{z}_{ijk}^{abc}$), right: $t_{ijk}^{abc}$ approximated with DF and RR ($\bar{t}_{ijk}^{abc}$); \textbf{(D)} Left: compressed $\bar{z}_{ijk}^{abc}$ ($\tilde{z}_{ijk}^{abc}$) , right: simplified representation; \textbf{(E)} Left: compressed $\bar{t}_{ijk}^{abc}$ ($\tilde{t}_{ijk}^{abc}$), right: simplified representation; \textbf{(F)} Alternative $\tilde{t}_{ijk}^{abc}$ formed using orthogonal rotation, right: simplified representation. 
    }
    \label{fig:tensor_dencom_def}
\end{figure}

\subsection{Density fitting of the ERIs}

Density fitting factorizes the four-dimensional ERI tensor into a contraction between two three-dimensional tensors via a resolution of the identity (RI) of an auxiliary basis set:\cite{Kendall1997, boys1959, Vahtras1993, Rendell1994}
\begin{align}
v^{pr}_{qs}&\approx\sum_{KL}(pq|K)(K|L)^{-1}(L|rs) \nonumber \\
&=\sum_{JKL}(pq|K)(K|J)^{-\frac{1}{2}}(J|L)^{-\frac{1}{2}}(L|rs) \nonumber \\
&=\sum_{J}B_{pq}^{J}B_{rs}^{J}
\label{eq:DF}
\end{align}
where $(pq|K)$ and $(K|L)$ are three- and two-center ERIs, and where $(K|L)^{-1}$ and $(K|L)^{-1/2}$ refer to specific matrix elements of the inverse and inverse square root of the total two-center ERI matrix. While the auxiliary basis can be constructed on-the-fly, pre-optimized auxiliary basis sets are generally the option of choice.\cite{Eichkorn1995, Bernholdt1998, Weigend2002_jcp, Weigend2002}  $B_{pq}^{J}$ are the density fitting factors, shown as yellow-green triangles in Fig.~\ref{fig:tensor_dencom_def}A. Testing on all the species in this work found that the use of DF-approximated ERIs introduced no more than 0.034 \kJmol of error in any of the (T) contributions to the reaction energies (see Supplemental Information) and thus is used as the ``reference'' calculation (Figure \ref{fig:Methods_proposed}B) for the analysis of the RR and Tucker approximations. A further reduction in the density fitting error can be achieved by using a larger auxiliary basis set, which has a relatively minor impact on the cost of the (T) correction, \emph{vide infra}.

\subsection{Rank-reduced $T_2$ amplitudes}

In rank-reduced CCSD\cite{Kinoshita2003, Parrish2019}, the $T_2$ amplitudes are approximated as,
\begin{align}
t_{ij}^{ab}\approx \sum_{VW} V_{ai}^{V}T^{VW}V_{bj}^{W}
\label{eq:RR_of_T2}
\end{align}
where $V_{ai}^{W}$ is an eigenvector of a reference set of $T_2$ amplitudes, obtained by folding into a $vo \times vo$ matrix and then diagonalizing. Typically these amplitudes are taken from an MP2 or MP3 calculation so that they may be fixed throughout the CCSD iterations. $T^{VW}$ is a core matrix that represents the compressed CCSD $T_2$ amplitudes. Alternatively, if one does not wish to accelerate CCSD, but rather to approximate (T), it is sufficient to perform an eigenvalue decomposition (EVD) of the converged CCSD $T_2$ amplitudes,
\begin{align}
t_{ij}^{ab}\approx \sum_{V} V_{ai}^{V} \lambda^{V}V_{bj}^{V}
\label{eq:EVD_of_T2_diagonal}
\end{align}
where $V_{ai}^{V}$ are the eigenvectors of the converged $T_2$ and $\lambda^{V}$ are the associated eigenvalues (diagrammatic representation in Figure \ref{fig:tensor_dencom_def}B). For this work, we adopt the more general formula in order to retain the option of using MP2 or MP3 $T_2$ compression vectors (or some other choice), in which case $T^{VW}$ is not diagonal. The decomposition is truncated using the absolute value of the $T_2$ eigenvalues as the metric of interest, and results in $N_{T_2} = \mathcal{O}(N)$ ``projectors'' retained in the compressed space. This decomposition incurs $\mathcal{O}(o^3 v^3)$ cost (or lower), and scales favorably compared to the $\mathcal{O}(o^2v^4)$ scaling of later steps in the process for most basis sets of interest. 

\subsection{Tucker-3 factorization of the $T_3$ amplitudes}

\begin{figure}[]
    \centering
    \includegraphics[width=0.35\textwidth]{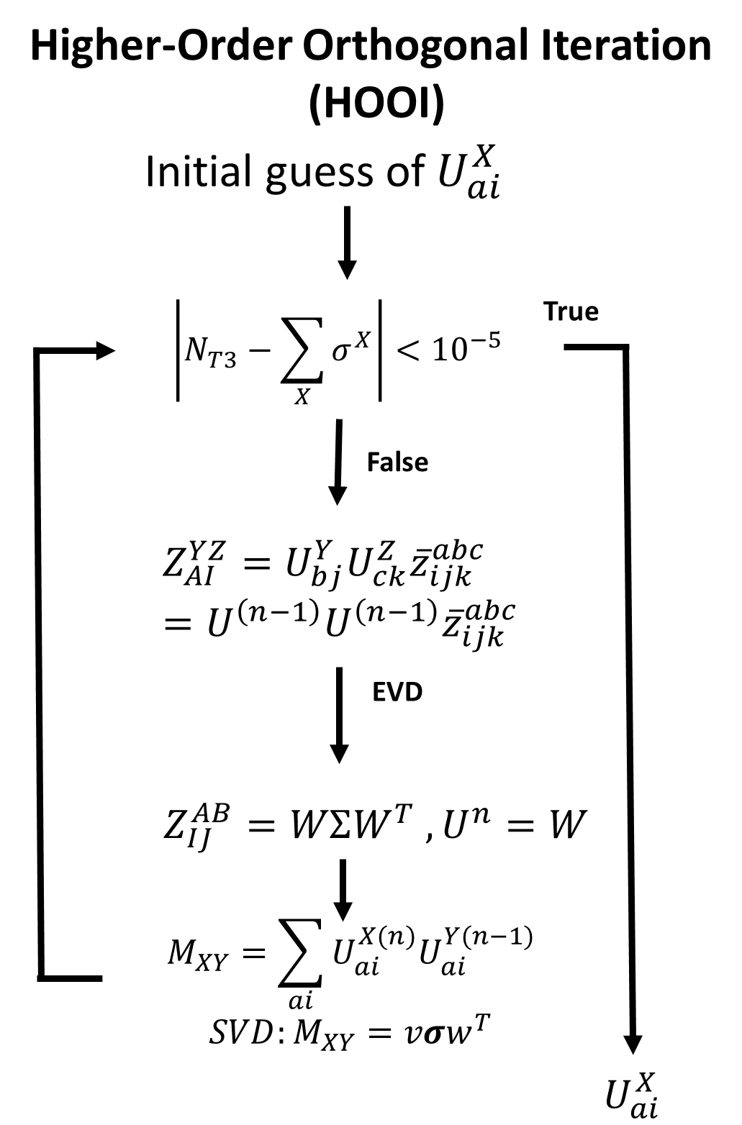}
    \caption{HOOI procedure demonstration diagram}
    \label{fig:HOOI}
\end{figure}

Starting with triples amplitudes defined using both the DF and RR approximations,
\begin{align}
\bar{z}_{ijk}^{abc}&=P_{ijk}^{abc}\left( \sum_{dJVW} B_{ck}^J B_{bd}^J V_{ai}^V T^{VW} V_{dj}^W -\sum_{lJVW} B_{ck}^J B_{jm}^J V_{ai}^V T^{VW} V_{bm}^W \right) \\
\bar{t}_{ijk}^{abc}&=\bar{z}_{ijk}^{abc}D_{ijk}^{abc}
\end{align}
the Tucker-3 compression of the $Z_3$ intermediates or $T_3$ amplitudes is defined as,
\begin{align}
\bar{z}_{ijk}^{abc}\approx  \tilde{z}_{ijk}^{abc}=\sum_{XYZ} Z^{XYZ}U_{ai}^{X}U_{bj}^{Y}U_{ck}^{Z} \\
\bar{t}_{ijk}^{abc}\approx  \tilde{t}_{ijk}^{abc}=\sum_{XYZ} T^{XYZ}U_{ai}^{X}U_{bj}^{Y}U_{ck}^{Z}
\label{eq:tucker}
\end{align}
where the graphical representations of each quantity are depicted in Fig.~\ref{fig:tensor_dencom_def}C--E. $Z^{XYZ}$ is the core tensor in the compressed basis and $U_{ai}^{X}$ are the compression vectors or projectors.\cite{Lesiuk2019, Lesiukacs2019} 
This representation can be obtained via a higher-order SVD (HOSVD) of a full, $T_3$-like quantity, but this process would be very expensive. Instead, following the work of Lesiuk\cite{Lesiuk2022} (and the earlier work of Bell, Lambrecht, and Head-Gordon in the context of MP2\cite{Bell-MP-2010}) we employ higher-order orthogonal iteration (HOOI) which constructs $U_{ai}^{X}$ in a self-consistent manner. In most of what follows, the initial guess for $U_{ai}^{X}$ is seeded from the eigenvectors of $T_2$. These are then partially contracted against the three-electron property in question, such as $\bar{z}^{abc}_{ijk}$, obtaining for example, ${Z}^{XY}_{ck}$,
\begin{align}
{Z}^{XY}_{ck}=\sum_{aibj}\bar{z}^{abc}_{ijk}U_{ai}^{X}U_{bj}^{Y}
\label{eq:tucker_(T)}
\end{align}
Because the projectors $U$ are orthogonal but rectangular, they form an approximate resolution of the identity. Thus, ${Z}^{XY}_{ck}$ contracted with itself to form a matrix $Z_{ck,c^\prime k^\prime}$, provides eigenvectors and eigenvalues approximating the right singular vectors and values of $\bar{z}^{abc}_{ijk}$ formatted as a matrix $\bar{z}_{aibj,ck}$,
\begin{align}
Z_{ck,c^\prime k^\prime} &= \sum_{XY}{Z}^{XY}_{ck} Z^{XY}_{c^\prime k^\prime} \nonumber \\
&= \sum_{\substack{aibj \\ a^\prime i^\prime b^\prime j^\prime }} \bar{z}^{abc}_{ijk} (\sum_X U^X_{ai} U^X_{a^\prime i^\prime }) (\sum_Y U^Y_{bj} U^Y_{b^\prime j^\prime }) \bar{z}^{a^\prime b^\prime c^\prime }_{i^\prime j^\prime k^\prime } \nonumber \\
&= \sum_{Z} U^{Z}_{ck} \epsilon^Z U^{Z}_{c^\prime k^\prime} \\
\bar{z}^{abc}_{ijk}&= \bar{z}_{aibj,ck} = \sum_Z L^Z_{aibj} \sigma^Z R^Z_{ck} \\
U^Z_{ck} &\approx R^Z_{ck},\quad \epsilon^Z \approx (\sigma^Z)^2
\label{eq:tucker_EVD}
\end{align}
where we assume a self-consistent solution of $U$. The positive square roots of $\epsilon^Z$ are used to determine the number of projectors ($N_{T_3}$) retained in the compressed space. The associated eigenvectors are then used as the next iteration of guesses for the Tucker-3 compression vectors. Convergence is checked by computing the singular values $\sigma^X$ of the overlap matrix $M_{XY}=\sum_{ai} U^{X(n)}_{ai} U^{Y(n-1)}_{ai}$ for projectors from successive iterations $n-1$ and $n$. If the value $|N_{T_3} - \sum_X \sigma^X|$ falls below some cutoff, chosen here as $10^{-5}$, then the process is stopped. A schematic of this procedure is shown in Fig.~\ref{fig:HOOI}. The final core tensor $Z^{XYZ}$ is then obtained by compressing $\bar{z}^{abc}_{ijk}$ with the converged $U$ factors,
\begin{align}
    Z^{XYZ}=\sum _{aibjck}\bar{z}_{ijk}^{abc}U_{ai}^{X}U_{bj}^{Y}U_{ck}^{Z}
\end{align}

\subsection{Laplace denominators and orthogonal rotation}

In order to avoid the construction of any full, three-electron tensor in our decomposition of $Z_3$ or $T_3$, we employ two approximations of the three-electron orbital energy denominators ($D_3$): Laplace transform (LT)\cite{Haser1992,Constans2000,constans2003,kats2008} and orthogonal rotation (OR)\cite{Hino2004, Lesiuk2022}. In LT, $D^{abc}_{ijk}$ is represented as a sum over Laplace transform quadrature points,
\begin{align}
\frac{1}{\varepsilon_{i}+\varepsilon_{j}+\varepsilon_{k}-\varepsilon_{a}-\varepsilon_{b}-\varepsilon_{c}}&=-\int_{0}^{\infty}e^{-(\varepsilon_{a}+\varepsilon_{b}+\varepsilon_{c}-\varepsilon_{i}-\varepsilon_{j}-\varepsilon_{k})t} \nonumber \\
&\approx - \sum_{g}\tau_{a}^{g}\tau_{b}^{g}\tau_{c}^{g}\tau_{i}^{g}\tau_{j}^{g}\tau_{k}^{g}
\end{align}
where $\tau_{i}^{g}=\omega_{g}^{1/6}e^{\varepsilon_{i}t_{g}}$ and $\tau_{a}^{g}=\omega_{g}^{1/6}e^{-\varepsilon_{a}t_{g}}$. $\omega_{g}$ and $t_{g}$ are the weights and abscissas of a suitable quadrature.\cite{braessApproximationExponentialSums2005} When used in the HOOI process described above, the partially-projected $T^{XY}_{ck}$ term is then constructed from a sum over these points,
\begin{align}
    T^{XY}_{ck} = -\sum_{g}T^{XY(g)}_{ck} \\
    T^{XY(g)}_{ck} = \sum_{aibj}\bar{z}^{abc}_{ijk}\tau^g_{a}\tau^g_{b}\tau^g_{c}\tau^g_{i}\tau^g_{j}\tau^g_{k}U^{X}_{ai}U^Y_{bj}
\end{align}
This Laplace transform technique has been known for some time and works remarkably well with between three and twelve Laplace points depending on the level of accuracy required and whether or not core orbitals are included.\cite{Haser1992,Constans2000,constans2003,kats2008}

Alternatively, one can perform an orthogonal rotation of the $U^X_{ai}$ projectors in order to simplify $D_3$ without the Laplace transform. When the denominators are applied to the compressed $\tilde{z}^{abc}_{ijk}$ intermediates, and then immediately recompressed, the projectors can be rotated to a ``diagonal" representation such that an equivalent set of orbital denominators can be applied directly to the core tensor $Z^{XYZ}$\cite{Hino2004},
\begin{align}
    T^{XYZ} &= D^{XYZ} Z^{XYZ} \\
    D^{XYZ} &= \frac{1}{\epsilon^X + \epsilon^Y + \epsilon^Z} \\
    U^{XY} &= \sum_{ai} U^X_{ai} (\varepsilon_i-\varepsilon_a) U^Y_{ai} = \delta_{XY} \varepsilon^X.
\end{align}
The required form can be achieved by diagonalizing the matrix $U^{XY}$ formed from the original projectors, and then rotating the projectors by the matrix of eigenvectors. Except where noted, a Tucker-3 compression of the $T_3$ amplitudes, denoted $\tilde{T}$ in Fig.~\ref{fig:tensor_dencom_def}, also implies the inclusion of $D_3$ during the HOOI process (that is, $T^{XY}_{ck}$ is used in place of $Z^{XY}_{ck}$), while compression of $Z_3$ uses $Z^{XY}_{ck}$. In the $T^{XY}_{ck}$ case, the Laplace denominators must be used during the HOOI.

\subsection{Approximation schemes}

\begin{table*}[]
    \centering
    \begin{ruledtabular}
    \begin{tabular}{c|r|r|r|r}
        \textbf{Method} &  \multicolumn{1}{c|}{\textbf{HOOI}} & \multicolumn{1}{c|}{$\mathbf{Z_{XYZ}}$} & \multicolumn{1}{c|}{$\mathbf{T_{ABC}}$} & \multicolumn{1}{c}{\textbf{Energy}} \\
    \hline
       $\Tilde{Z}T$ & $N_{T_3}^2v^2o^2*N_{\text{it}}$ & $N_{DF}N_{T_3}^{2}vo$& 0&$N_{DF}N_{T_3}^{2}vo^2*N_{g}$ \\ 
       
        $\Tilde{T}Z$ & $N_{T_3}^2v^2o^2*N_{g}*N_{\text{it}}$ & 0 &$N_{DF}N_{T_3}^{2}vo*N_{g}$& $N_{DF}N_{T_3}^{2}vo^2$\\
        
        $\Tilde{Z}^\prime\Tilde{D}Z$ & $N_{T_3}^2v^2o^2*N_{g}*N_{\text{it}}$ &0 & $N_{DF}N_{T_3}^{2}vo$& $N_{DF}N_{T_3}^{2}vo^2$ \\ 
        
        $\Tilde{Z}\Tilde{D}Z$ & $N_{T_3}^2v^2o^2*N_{\text{it}}$ & 0 &$N_{DF}N_{T_3}^{2}vo$ & $N_{DF}N_{T_3}^{2}vo^2$\\
        
        $\Tilde{Z}\Tilde{D}\Tilde{Z}$ & $N_{T_3}^2v^2o^2*N_{\text{it}}$& $N_{DF}N_{T_3}^{2}vo$  & $N_{T_3}^3$ & $N_{T_3}^{3}vo^{2}$\\
 
    \end{tabular}
    \end{ruledtabular}
    \caption{Scaling of the leading term contributing to distinct phases of the approximations discussed. $N_{\text{it}}$ is the number of iteration cycles in the HOOI procedure.}
    \label{tab:methods_scaling}
\end{table*}

\begin{figure}[]
    \centering
    \includegraphics[width=0.5\textwidth]{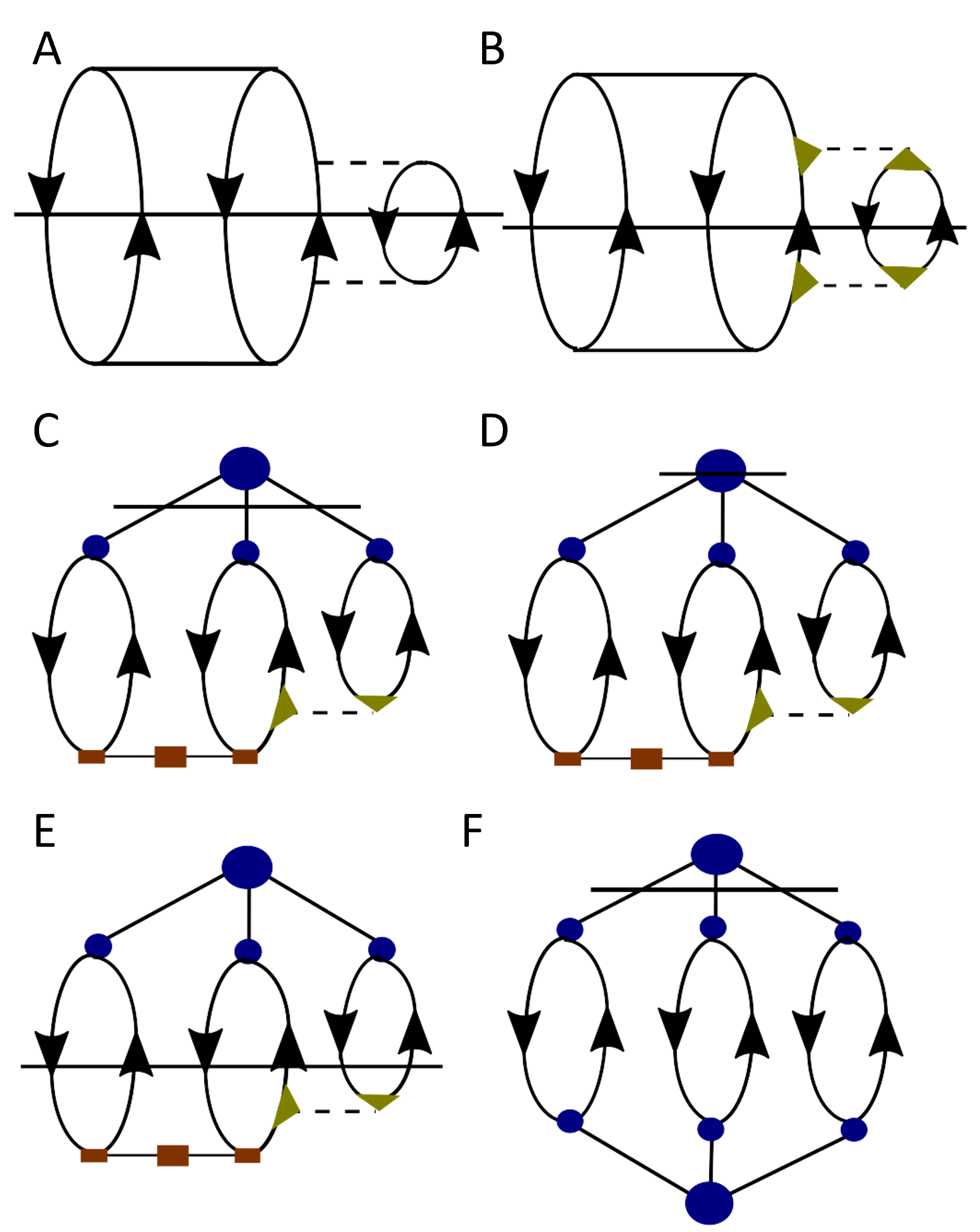}
    \caption{Methods explored in this work. \textbf{(A)} Canonical (T);\textbf{(B)} Canonical (T) with density fitting; \textbf{(C)} $\Tilde{Z}\Tilde{D}Z$ or $\Tilde{Z}^{\prime}\Tilde{D}Z$; \textbf{(D)} $\Tilde{T}Z$; \textbf{(E)}  $\Tilde{Z}T$; \textbf{(F)} $\Tilde{Z}\Tilde{D}\Tilde{Z}$.}
    \label{fig:Methods_proposed}
\end{figure}

We now use $E^{[4]}$ to demonstrate how the various approximations discussed above can create a number of reduced scaling implementations of CCSD(T). The most important term to approximate is either $z_{ijk}^{abc}$ or its denominator-weighted version, ${t}_{ijk}^{abc}$. These terms constitute the two halves of the $E^{[4]}$ diagram in Fig.~\ref{fig:Methods_proposed}A. 
In order to guarantee $\mathcal{O}(N^6)$ scaling during the HOOI and in the calculation of the energy for all methods, density fitting is applied to the Hamiltonian term and the rank-reduced compression is applied to $T_2$ (Figure \ref{fig:tensor_dencom_def}C). Then, one or both of $\bar{z}_{ijk}^{abc}$ and $\bar{t}_{ijk}^{abc}$ are cast into Tucker-3 form ($\tilde{z}_{ijk}^{abc}$ and $\tilde{t}_{ijk}^{abc}$), creating five distinct reduced scaling methods:

\begin{itemize}
    \item{{$\mathbf{\tilde{Z}\tilde{D}Z}$} The top $Z_3$ is Tucker-3 compressed to form $\tilde{Z}_3$, which is then transformed into $\tilde{T}_3$ via orthogonal rotation. The bottom $Z_3$ remains uncompressed. See Figure \ref{fig:Methods_proposed}C.}

    \item{{$\mathbf{\tilde{Z}^\prime \tilde{D} Z}$} As in $\tilde{Z}\tilde{D}Z$, except that $T^{XY}_{ck}$ is used during the HOOI instead of $Z^{XY}_{ck}$.}

    \item{{$\mathbf{\tilde{T}Z}$} The top $Z_3 D_3$ is Tucker-3 compressed into $\tilde{T}_3$, while the bottom remains as $Z_3$. This is the method previously published by Lesiuk.\cite{Lesiuk2022} See figure \ref{fig:Methods_proposed}D.}

    \item{{$\mathbf{\tilde{Z}T}$} The top $Z_3$ is Tucker-3 compressed into $\tilde{Z}_3$ and the bottom $Z_3 D_3 = T_3$ is left uncompressed. Note that requires application of the Laplace denominators during evaluation of the energy. See Figure \ref{fig:Methods_proposed}E.} 

    \item{{$\mathbf{\tilde{Z}\tilde{D}\tilde{Z}}$} The top and bottom $Z_3$ are Tucker-3 compressed to form $\tilde{Z}_3$. The top $\tilde{Z}_3$ is then transformed into $\tilde{T}_3$ via orthogonal rotation. See Figure \ref{fig:Methods_proposed}F.}
    
\end{itemize}
It should be noted that there are additional possible schemes utilizing the same approximations, however initial screening clearly singled out these five methods as the most accurate; see Section~\ref{sec:results} for more discussion on this point.

The scaling of the leading order terms for each of these five reduced scaling methods are displayed in Table~\ref{tab:methods_scaling}, and the full analysis is provided in the Supplementary Information. Note especially that the scaling of the HOOI term can in general be lowered to $\mathcal{O}(N^5)$ if the rank-reduced $T_2$ amplitudes are fully leveraged. In order to test un-approximated $T_2$ as well, we keep a slightly lower-performing factorization. Additionally, the use of randomized projection techniques\cite{Jiang-JCTC-2023} can further accelerate this portion of the calculation. For each method, we separately report the leading order term in a) the HOOI, b) the compression of $Z_3$ (if done), c) the compression of $T_3$ (if done), and d) the evaluation of the approximate (T) energy correction. All five methods achieve $\mathcal{O}(N^6)$ scaling, allowing that $N_{\text{it}}$ and $N_{\text{g}}$ are essentially constant. There are three distinct types of methods proposed here: those that employ the orthogonal rotation technique in order to incorporate the orbital energy denominators (although the Laplace denominators are required in the HOOI step of $\tilde{Z}^\prime \tilde{D} Z$), those that directly employ the Laplace denominators in either the compression of $T_3$ or the evaluation of the energy, and, by itself, the $\tilde{Z}\tilde{D}\tilde{Z}$ method which employs a double compression. Note that the similar method $\tilde{T}\tilde{Z}$ is not presented here, for reasons which will become clear in Section~\ref{sec:results}.

\section{Computational details}

The accuracy of the above approximation schemes was tested on a subset of the W4-17 data set\cite{Karton2017} detailed in Table~\ref{tab:species}. The species were selected to represent a range of bonding environments, number of heavy atoms, and multi-reference character. The geometries are directly obtained from the W4-17 supporting information. Following the W4 protocol, aug-cc-pV$X$Z\cite{dunningbasis,Kendall1992} and aug-cc-pCV$X$Z\cite{dunningbasis,Kendall1992,Woon1995} basis sets were used on non-hydrogen atom valance and all-electron CCSD(T) correlation energies, respectively, while cc-pV$X$Z basis sets were employed on hydrogen atoms. RI basis sets\cite{Feyereisen1993,dunlapdf,auxbasis} of the same were used to construct the auxiliary basis for density fitting. Basis-set extrapolations of the CCSD(T) correlation energy were obtained using the two-point extrapolation formula using the basis set cardinality $X$ ($X = 3, 4, 5$ for TZ, QZ, 5Z, etc.),
\begin{align}
    E(X) = E_{\infty} + A(X+a)^{-\alpha} 
\end{align}
where $a$ and $\alpha$ are taken as 0.5 and 4, respectively.\cite{Martin1996, Feller2011, Feller2013} There are other extrapolation schemes that may be chosen, but this will not significantly change the relative performance of the methods explored here. Extrapolated values are indicated in the $\{X-1, X\}$ notation such that aug-cc-pV$\{T,Q\}$Z would indicate a reaction energy or correlation energy obtained from the extrapolation using aug-cc-pVTZ/cc-pVTZ (non-hydrogen/hydrogen) and aug-cc-pVQZ/cc-pVQZ (non-hydrogen/hydrogen) basis sets.

Aside from density fitting (controlled by the auxiliary basis set used as noted above) and the Laplace denominators (controlled by the number of quadrature points, which is 12 in all presented calculations), the cost and accuracy of the approximate methods are controlled by two adjustable parameters, $N_{T_2}$ and $N_{T_3}$, which determine the size of the rank-reduced and Tucker-3 compressions of $T_2$ and $T_3$. We report these values as multiples of the number of basis functions $N_{bas}$. The HOOI procedure was terminated when $ |N_{T_3} - \sum_X \sigma^X| < 10^{-5}$, which usually occurs in six iterations. All calculations were performed in a development version of {\sc cfour}.\cite{cfour} 

The ANL$n$ reaction scheme\cite{Klippenstein2017} was used to benchmark the accuracy of the above approximate methods against the canonical DF-CCSD(T) reaction energies using the same basis sets. This scheme was extended to include boron, silicon, sulphur, and chlorine-containing species. \ce{H2}, \ce{B2H6}, \ce{CH4}, \ce{H2O}, \ce{NH3}, \ce{HF}, \ce{HCl}, \ce{H2S}, and \ce{SiH4} were considered as reference species for the H, B, C, O, N, F, Cl, S, and Si elements, respectively, such that the appropriate reaction products of an arbitrary molecule can be determined as
\begin{align}
  \ce{H_a B_b C_c N_d O_e F_f Si_g S_h Cl_i} &= x\ce{H2} + b\ce{B2H6} + c\ce{CH4} + d\ce{NH3} \nonumber\\ 
  &+ e\ce{H2O} + f\ce{HF} + g\ce{SiH4} \nonumber\\
  &+ h\ce{H2S} + i\ce{HCl},
\end{align}
where the number of \ce{H2} molecules in the products, $x$, needed to balance the equation can be determined algebraically. While this reaction scheme is not designed to universally take advantage of error cancellation (for example, the carbon dioxide reaction scheme would be \ce{4H2 + CO2 -> 2H2O + CH4}, which preserves no bonding character between the reactants and products), all the reference species involved are closed shell and free from significant multireference effects. 

\begin{table}[]
    \centering
    \begin{ruledtabular}
    \begin{tabular}{lllllll}
    \multicolumn{1}{l}{\#  of Heavy}    & \multicolumn{2}{c}{Single Reference}  & \multicolumn{2}{c}{Minor MR}  & \multicolumn{2}{c}{ Moderate MR}  \\
    \multicolumn{1}{l}{Atoms }  & \multicolumn{2}{c}{($\%TAE[(T)]<4$)} &   \multicolumn{2}{c}{($4<\%TAE[(T)]<10$)} &        \multicolumn{2}{c}{($\%TAE[(T)]>10$)}    \\ \hline
    \multirow{2}{*}{2}               & \multicolumn{1}{l}{\ce{B2H6}} & 0.74 & \multicolumn{1}{l}{\ce{N2}} &      4.16  & \multicolumn{1}{l}{\ce{B2}} &  14.66  \\
                                     & \multicolumn{1}{l}{\ce{SiH3F}} &  0.81 & \multicolumn{1}{l}{\ce{O2}} &     7.67   & \multicolumn{1}{l}{\ce{C2}} & 13.26    \\
    \hline
 \multirow{4}{*}{3}& \ce{BF2H} & 1.49 & \ce{HCNO}   &4.77& \ce{O3} & 17.39\\
 & \ce{CH3CHO} &1.72& \ce{HN3} &5.64&\ce{F2O} & 14.6\\
 & \ce{C3H4} &1.75&  && &\\
 & \ce{C2H5F} &1.25&  && &\\  
    \hline
 4& \ce{BF3} &1.77& \ce{SO3} &5.87&\ce{F2O2} &16.92\\   
 5& \ce{SiF4} &1.75& \ce{HClO4}  &7.92&  &\\   
 6& \ce{C6H6} &1.96& \ce{N2O4} &9.11&\ce{ClF5} &14.78\\  
 7&  && \ce{SF6} &4.29& &\\   
    \end{tabular}
      \end{ruledtabular}
    \caption{Molecule set selected from W4-17, along with $\%TAE[(T)]$ values from Ref.~\citenum{Karton2017} which approximately indicate multi-reference (MR) behavior.}
    \label{tab:species}
\end{table}

\begin{sidewaysfigure}[]
    \centering
    \includegraphics[width=1.0\textwidth]{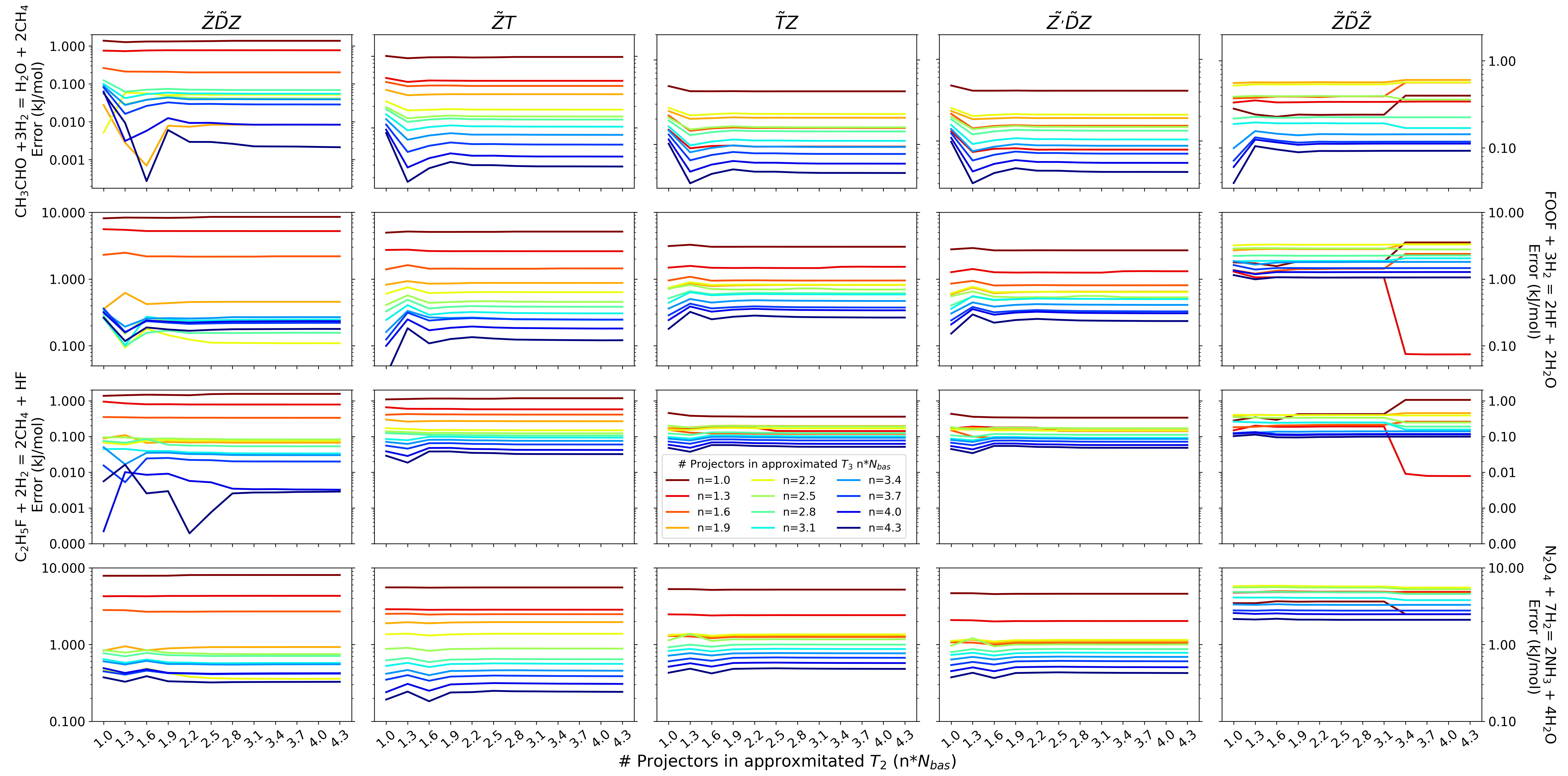}
    \caption{$T_2$ convergence and $T_3$ convergence behavior for different methods and four representative reactions. The absolute reaction energy error with the aug-cc-pV\{T,Q\}Z extrapolation is shown on the y-axis. The x-axis indicates the number of $T_2$ projectors as a multiple $n$ ($N_{T_2}=nN_{bas}$). Separate colored lines correspond to increasing (red to blue) numbers of $T_3$ projectors ($N_{T_3}=mN_{bas}$).}
    \label{fig:t2_convergence}
\end{sidewaysfigure}

\section{Results and Discussion}\label{sec:results}

\subsection{Separability of the approximations}

Each of the above approximate schemes is constructed from individual compressions or approximations of the two-electron ERIs (only density fitting is used here), the $T_2$ amplitudes (rank-reduced compression with a user-defined $N_{T_2}$), and Tucker-3 compression of the $T_3$ amplitude or $Z_3$ intermediates (with a user-defined $N_{T_3}$ and convergence threshold of the HOOI procedure). As might be expected, density fitting of the ERIs introduces exceptionally small errors (less than 0.5\% of the (T) correction) for the reaction energies studied here and is applied to all ERI terms in the approximate (T) schemes and in the CCSD calculation which determines $T_2$. More information on the effect of density fitting in the present calculations can be found in the Supporting Information. The HOOI procedure, the only iterative portion of these approximation schemes, converges rapidly 
to meet a very tight tolerance of $10^{-5}$. Further, it appears that the predicted reaction energies depend only minimally upon the number of HOOI iterations; even a loose convergence threshold (or potentially, a non-iterative determination of $U$) appears to be sufficient to achieve convergence in the reaction energies (see Fig.~S1). Thus, consideration of the DF and HOOI treatments can be separated from the $T_2$ rank-reduction and $Z_3/T_3$ compression. These findings are consistent with those observed in previous literature.\cite{Lesiuk2022}

At first glance, it is not obvious that this separability should extend to the compression of the $T_2$ and $Z_3/T_3$ space, especially considering that the compressed $T_2$ amplitudes are used in the construction of the compressed three-electron quantities. Fig.~\ref{fig:t2_convergence} displays the dependence of a representative subset of reaction energies on the number of $T_2$ projectors, with separate curves accounting for increasing numbers of $T_3$ projectors. Increasing the number of $T_2$ projectors past roughly $2.2N_{bas}$ leads to no significant improvement in the predicted reaction energies for all numbers of $T_3$ projectors. Additionally, the curves for separate numbers of $T_3$ projectors run essentially parallel across the $N_{T_2}$ space, indicating that these two approximations are decidedly separable. Thus, we fix the number of $T_2$ projectors to a ``safe" value of $3N_{bas}$ for all reactions, effectively making the number of $T_3$ projectors the sole variable that controls both cost and accuracy of the five approximate methods studied. Note that $N_{T_2}$ has minimal impact on the cost of the overall energy evaluation.

\begin{figure*}[]
    \centering
    \includegraphics[width=1.0\textwidth]{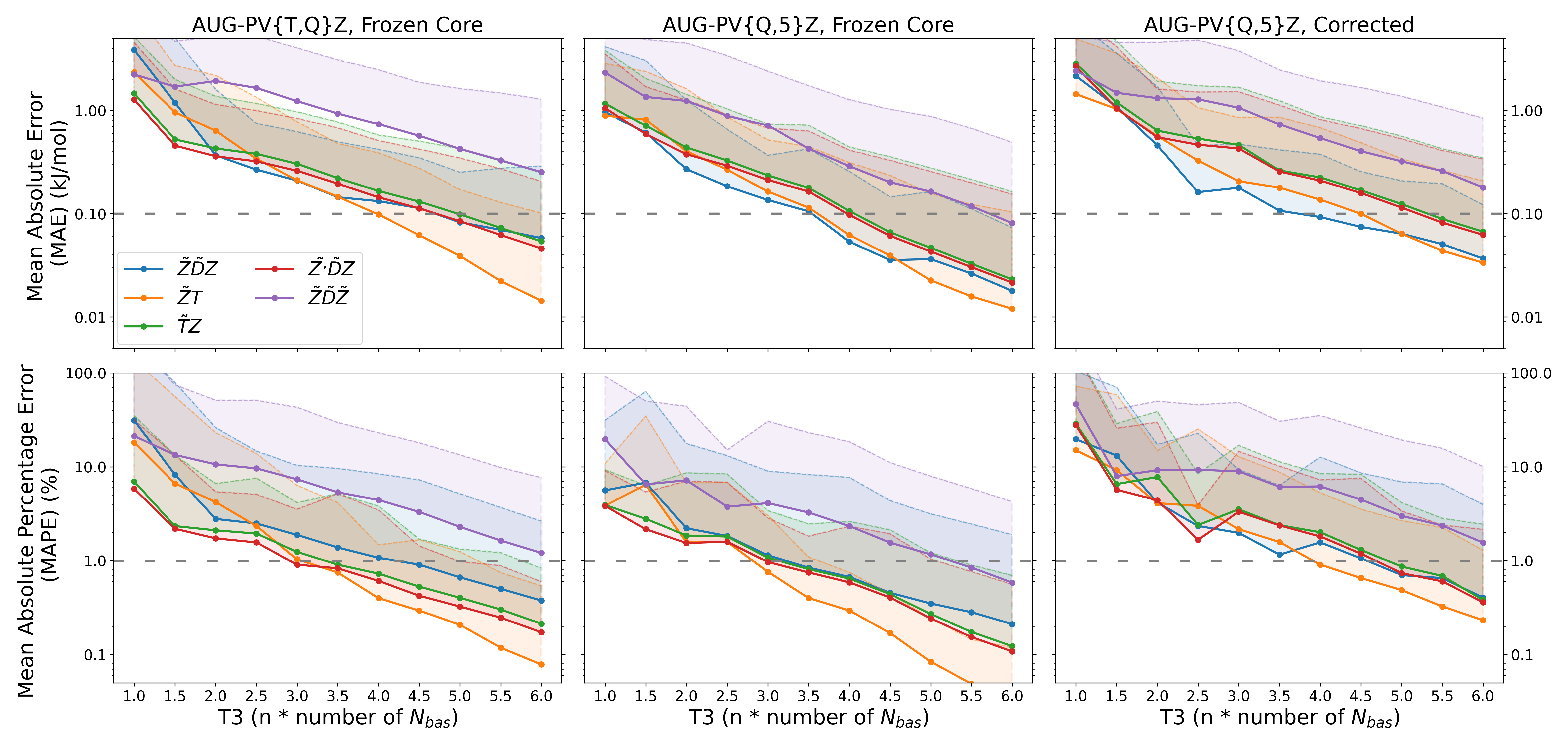}
    \caption{Error statistics of various decomposition schemes for the benchmark reaction set. The thick lines indicate the mean absolute or absolute percentage errors across the set, while the top of the shaded region indicates the maximum error. Plots are labeled by the extrapolation scheme used, and ``aug-cc-pV\{Q,5\}Z Corrected" indicates a frozen core aug-cc-pV\{Q,5\}Z valance-only energy plus an aug-cc-pCV\{T,Q\}Z core-valance correction. \ce{HClO4}, \ce{C6H6}, \ce{SF6} and \ce{ClF5} are excluded from the aug-cc-pV\{Q,5\}Z test set. }
    \label{fig:stat_mean_all}
\end{figure*}

\begin{figure*}[]
    \centering
    \includegraphics[width=1.0\textwidth]{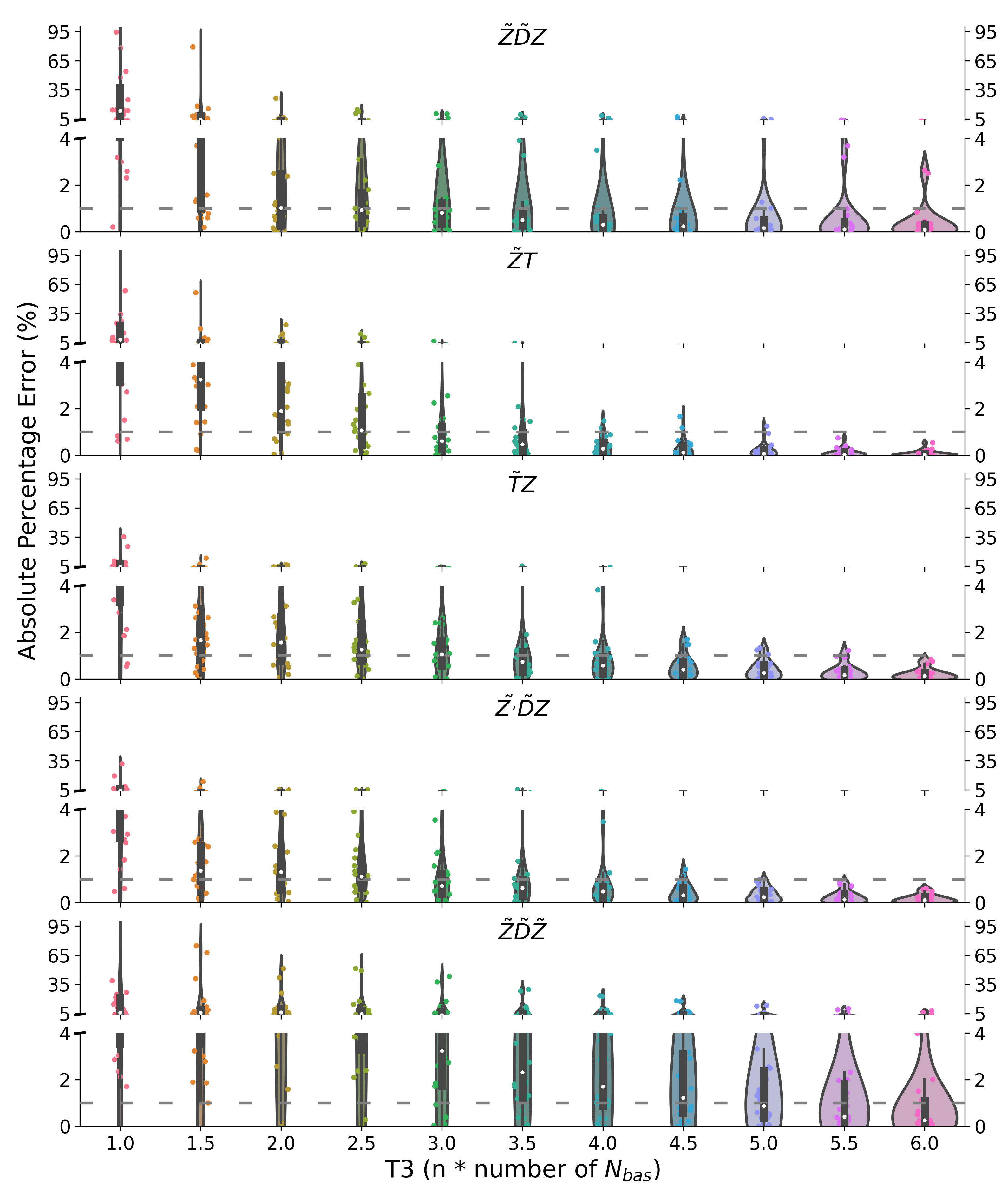}
    \caption{``Violin plots" of the mean absolute percent errors (MAPE) of reaction energies with the aug-cc-pV\{T,Q\}Z basis set extrapolation. The break in the y-axis is used to present the extremal data in detail.}
    \label{fig:stat_vilion}
\end{figure*}

\subsection{Comparison of approximation schemes in frozen core and core-valence corrected calculations}

To recapitulate, the purpose of this work is to assess the viability of several approximate (T) methods in the context of sub-chemical (1 \kJmol) accuracy reaction energies. Given that the perturbative triples are but one component in most model chemistries, we have arbitrarily selected 0.1 \kJmol as the target mean absolute error (MAE) and 1\% as the target mean absolute percent error (MAPE) for these approximate schemes applied to the test-suite of reaction energies. The number of $Z_3/T_3$ projectors required to hit these goals then determines the relative cost of these methods, though these targets may be changed to suit the needs of the individual model chemistry that these methods would be employed. 

Figure \ref{fig:stat_mean_all} displays the mean absolute error (MAE) and mean absolute percent error (MAPE) for the five approximate schemes discussed above for valence aug-cc-pV\{T,Q\}Z, valence aug-cc-pV\{Q,5\}Z, and core-valance corrected aug-cc-pV\{Q,5\}Z calculations, except that the largest systems \ce{HClO4}, \ce{C6H6}, \ce{SF6} and \ce{ClF5} are excluded from the aug-cc-pV\{Q,5\}Z test set. In this figure, the solid line indicates the mean value for the respective error statistic, while the upper edge of the shaded region represents the maximum error across the test set. All methods are capable of achieving a sub-0.1 \kJmol{}  and/or sub-1\% average accuracy given a large enough set of $Z_3/T_3$ projectors, though with varying error profiles. The $\tilde{Z}\tilde{D}\tilde{Z}$ scheme (purple line) requires the largest number of projectors to reach target MAE accuracy, which is perhaps unsurprising as this scheme employs Tucker-3 decompositions of both the top and bottom half of the $E^{[4]}$ diagrams. $\tilde{T}Z$ and $\tilde{Z}^\prime \tilde{D} Z$ (green and red lines, respectively) performed quite similarly, not just for the statistical errors, but also in the individual reaction energies, though $\tilde{Z}^\prime \tilde{D} Z$ (red line) is slightly more accurate. These two methods both create a Tucker-3 representation of $T_3$, but in $\tilde{Z}^\prime \tilde{D}Z$ the effect of the denominators is included via orthogonal rotation rather than the Laplace transform (although LT is still used in the HOOI iterations). The main differentiating factor between those potential approximation schemes reported and not reported here is the handling of the denominators in the determination of $U$ via HOOI. For example, $\tilde{T}Z$ includes the denominators in the HOOI since it is the $T_3$ amplitudes which are being directly approximated. The similar method $\tilde{T}^\prime Z$ which differs only in that it \emph{does not} include the denominators during the HOOI is much less accurate. Similar comparisons can be made for almost every method tested. This indicates a strong dependence of the projectors on the structure of $Z_3$ vs. $T_3$. In this light, the $\tilde{Z}\tilde{D}Z$ method (and to some extent, $\tilde{Z}\tilde{D}\tilde{Z}$) is then an outlier because it only slightly underperforms the ``correct" version $\tilde{Z}^\prime\tilde{D}Z$ (note that $\tilde{Z}\tilde{D} \sim \tilde{T}$, indicating that including denominators in the HOOI should perform better). There is also an inconsistency for $\tilde{Z}^\prime \tilde{D}Z$ compared to $\tilde{T}Z$---the former method is identical to the latter if the initial compression of $Z_3$ is removed (orthogonal rotation is an exact representation of the denominators in the compressed space so these would revert to the similarly ``exact" Laplace denominators). Instead, $\tilde{Z}\tilde{D}Z$ often outperforms $\tilde{T}Z$, particularly when the core-valence correlation is included. This may indicate a partial cancellation of errors between the compression of $Z_3$ and later $T_3$. 
 In the future, we will explore the use of multiple sets of projectors to further improve accuracy.
$\tilde{Z}T$ seems to perform best or second-best for the MAE and MAPE measures (and for the maximum absolute/percentage errors), while the relative performance of $\tilde{Z}\tilde{D}Z$ changes depending on the basis sets involved. This in turn may indicate a slightly better compressibility of $Z_3$ compared to $T_3$.

\begin{figure}[]
  \centering
  \includegraphics[width=0.92\textwidth]{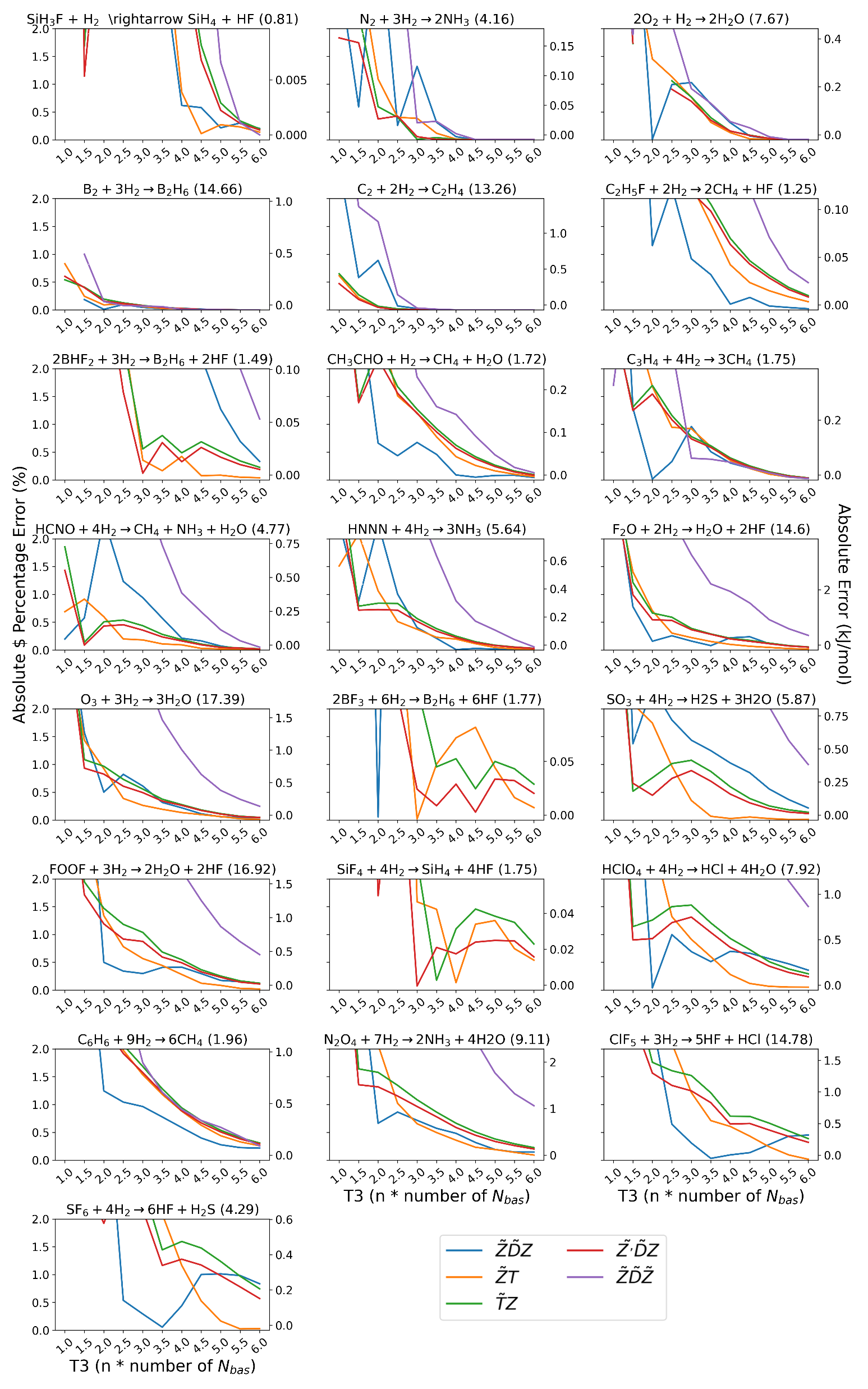}
   \label{fig:t3_each_species}
\end{figure}
\begin{figure}[t]
  \caption{Absolute relative errors (in percent) of each reaction species at the aug-cc-pV\{T,Q\}Z frozen-core level. In each figure, the right axis gives the absolute error in \kJmol.}
\end{figure}

Figure \ref{fig:stat_vilion} displays violin plots of the MAPE for each of these methods. Consistent with the average and max error plots,  $\tilde{T}Z$, $\tilde{Z}^\prime\tilde{D}Z$, and $\tilde{Z}T$ are most resistant to outliers. $\tilde{Z}T$ in particular obtains a tight distribution of errors compared to the other methods. $\tilde{Z}\tilde{D}\tilde{Z}$, the double-Tucker-3 method, performs worst, and $\tilde{Z}\tilde{D}Z$ performs well with the exception of a few outliers that skew the average statistics. These outliers correspond to the reactions : \ce{2BF3 + 6H2 -> B2H6 + 6HF}, \ce{SiF4 + 4H2 -> SiH4 4HF}, and \ce{2BHF2 + 3H2 -> B2H6 + 2HF}. Interestingly, these reactions all display extremely small (T) contributions to reaction energy. Thus, even large percentage errors still meet the desired \SI{0.1}{\kilo\joule\per\mol} threshold for these reactions. 

Most important in the potential application of these methods in model chemistries is the number of $Z_3/T_3$ projectors (which determines the overall computation cost) required to achieve some target accuracy. For example, a relatively expensive approximation scheme such as $\tilde{T}Z$ (green) may out-perform a less expensive scheme such as $\tilde{Z}T$ (orange) if the former requires a sufficiently smaller number of projectors to obtain equivalent confidence intervals. To explore this, we examine the number of projectors required to obtain an MAE of 0.1 \kJmol{} and estimate the relative expense of the five methods using the equations contained within the Supplementary Information. From this perspective, the two most promising methods are $\tilde{Z}T$ (orange) and $\tilde{Z}\tilde{D}Z$ (blue). $\tilde{Z}\tilde{D}\tilde{Z}$ (purple) falls behind as it requires a large number of $T_3$ projectors to obtain target accuracy (and scales cubically with $N_{T_3}$ while the other methods scale quadratically), while $\tilde{T}Z$ (green) and $\tilde{Z}^\prime\tilde{D}Z$ (red), although relatively stable, do not out-compete $\tilde{Z}T$'s (orange) ratio of cost to accuracy. Note that, in our present factorization of the working equations and implementation, the cost of the HOOI is a major factor in the overall wall time. Improvements to the HOOI (fewer iterations, random projection, etc.) could change the relative ranking of the methods. While a full analysis of performance data and empirical scaling is beyond the scope of this work, we do note already significant speedups compared to DF-CCSD(T) even for the present test set of small molecules.

It should be noted that while the accumulated statistics displayed in Fig.~\ref{fig:stat_mean_all} are used in the above discussion (and demonstrate a roughly linear profile in the log space of the respective errors), the errors in each individual reaction (see Fig.~\ref{fig:t3_each_species}) are frequently non-monotonic, and the conclusions above should therefore be taken as general commentary on possible applicability rather than guidance on constructing model chemistry using these techniques. We are currently working on further characterizing the errors in these methods and how to best predict the number of $T_3$ projectors required to obtain a target accuracy, but these are the subject of future manuscripts.

\begin{figure}[]
    \centering
    \includegraphics[width=\textwidth]{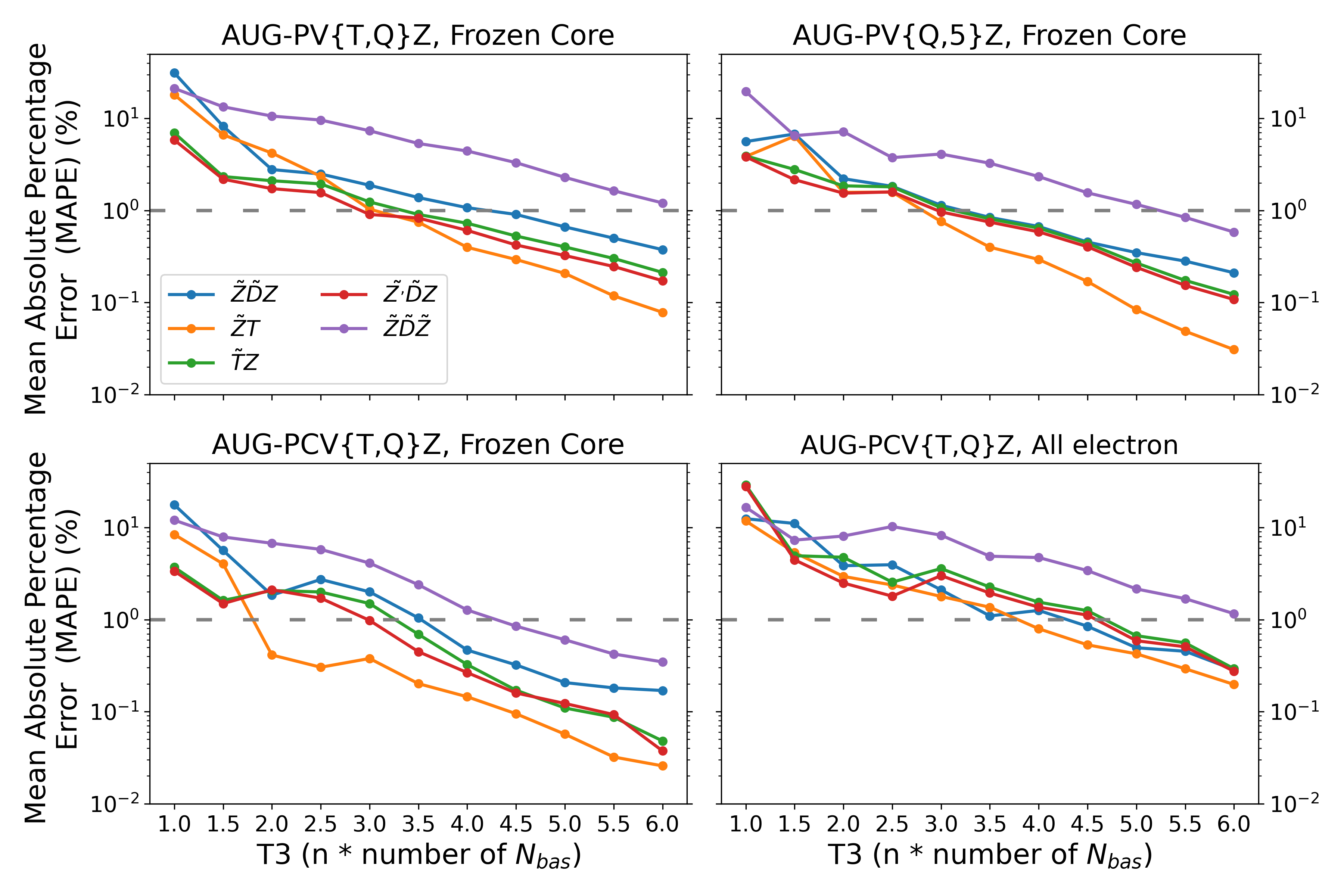}
    \caption{Mean absolute percent errors (MAPE) of extrapolated reaction energies with different combinations of basis sets and core-valence correlation. \ce{HClO4}, \ce{C6H6}, \ce{SF6} and \ce{ClF5} are excluded from these results, except for aug-cc-pCV\{T,Q\}Z frozen core.}
    \label{fig:all_electron}
\end{figure}

\subsection{Comparison of approximation schemes in all-electron calculations}

All high-accuracy model chemistries employ some variety of core-valance correlation corrections, typically calculated with aug-cc-pCV$X$Z variants of smaller canonical order than the aug-cc-pV$X$Z basis sets used to capture valance correlation. The exception to this rule is HEAT-like model chemistries\cite{Tajti2004, Bomble2006, Harding2008, Thorpe-JCP-2021}, which avoid potential additivity problems with this differing basis set treatment by correlating all electrons through CCSD(T). Fig.~\ref{fig:all_electron} displays the MAE statistics for the five above approximate schemes for aug-cc-pV\{T,Q\}Z frozen-core, aug-cc-pV\{Q,5\}Z frozen-core, aug-cc-pCV\{T,Q\}Z frozen-core, and aug-cc-pCV\{T,Q\}Z all-electron calculations. \ce{HClO4}, \ce{C6H6}, \ce{SF6} and \ce{ClF5} are excluded from these results, except for aug-cc-pCV\{T,Q\}Z with a frozen core. As expected, all the methods require a smaller ratio of $Z_3/T_3$ projectors to total basis set size when increasing the size of the basis sets used in valance calculations from aug-cc-pV\{T,Q\}Z to aug-cc-pCV\{Q,5\}Z. It has been previously shown that there are only a linear number of projectors that are important in describing molecular systems\cite{Parrish2019}, and thus, these methods become comparatively less expensive with larger basis sets where they can project onto the parts of correlation that ``matter'' from a larger space of basis functions. This is a very desirable property from the perspective of model chemistry development: the larger the basis set, the better, relatively, these rank-reduced techniques perform. 

However, while all methods require comparatively more projectors when treating all-electron correlation energies (which causes the upward shift in the aug-cc-pV\{Q,5\}Z core-valence corrected plots in Figure \ref{fig:stat_mean_all}), not all of the five approximate methods discussed here respond equivalently to the inclusion of core-electrons in the correlated wavefunction. In particular, the $\tilde{Z}T$ method, which generally performs best in valance-only aug-cc-pV\{T,Q\}Z, performs only a little better than its competitors for all-electron aug-cc-pCV\{T,Q\}Z. The underlying cause for this change is unclear and warrants further investigation in future work. However, there seems to be no reason to believe that these techniques cannot be used to address core-valance correlation in model chemistries.

\subsection{Discussion}

There are some interesting conclusions one can draw by making comparisons of the relative performance of these five approximate (T) methods that may provide insight for future schemes that use these techniques. 

$\tilde{Z}\tilde{D}\tilde{Z}$ is unequivocally the least accurate method studied here, likely due to the use of two Tucker-3 compressions. As shown above, the number of $T_3$ projectors is the primary determinant of the accuracy of these methods, and $\tilde{Z}\tilde{D}\tilde{Z}$ trades off decreased scaling with the number of density-fitting auxiliary basis functions for an increased scaling with $N_{T_3}$. Given that it requires more $T_3$ projectors than DF auxiliary functions to obtain the accuracy goals desired here, this trade-off simply doesn't net any gain.

As stated above, $\tilde{Z}^\prime\tilde{D}Z$ and $\tilde{T}Z$ differ only in that the former performs an orthogonal rotation of $\tilde{T}_3$ to compress the denominator terms while the latter uses a Laplace transform and quadrature. As the orthogonal rotation is exact within the compressed subspace (and the additional compression of $Z_3$ does not seem to lead to additional error) and the Laplace transform is a very precise approximation, there are only slight measurable differences between these methods even on a reaction-by-reaction basis, and the only discrepancy is the cost of performing the orthogonal rotation vs the cost of utilizing the Laplace transform. Both methods perform a Tucker-3 compression of the amplitudes rather than the residuals, and, in this case, both are outperformed by the $\tilde{Z}T$ method on a cost basis due to an increased cost of the HOOI step. 

Likewise, $\tilde{Z}\tilde{D}Z$ and $\tilde{Z}T$ both perform a Tucker-3 compression of the $Z_3$ intermediates and differ only in how they compress the denominator, as with $\tilde{Z}^\prime\tilde{D}Z$ and $\tilde{T}Z$. Unlike the above case, however, this results in significant differences between the two methods, especially when considered on a reaction-by-reaction basis. Generally it seems that $\tilde{Z}T$ is a bit more stable: it performs on par or better than $\tilde{Z}^\prime\tilde{D}Z$ and $\tilde{T}Z$ with a significantly reduced cost in the HOOI iterations, and only performs significantly worse than these in the \ce{2BF3 + 6H2 -> B2H6 + 6HF} reaction, for which the reaction energy is very small. $\tilde{Z}\tilde{D}Z$, on the other hand, is somewhat more erratic---an undesirable trait in an approximation to be used in high-accuracy thermochemistry.

The final comparison we make is between $\tilde{Z}^\prime\tilde{D}Z$ and $\tilde{Z}\tilde{D}Z$. The only difference between these methods is whether or not the Laplace denominators are included in the HOOI, which determines the projectors $U$. Both then undergo orthogonal rotation to create the $\tilde{T}_3$ used in the final energy calculations. However, while $\tilde{Z}^\prime\tilde{D}Z$ performs extremely similarly to its Laplace transform counterpart ($\tilde{T}Z$), $\tilde{Z}\tilde{D}Z$ features qualitatively different behavior to $\tilde{Z}T$, especially on a species-by-species comparison. In the case of $\tilde{Z}^\prime\tilde{D}Z$, we noted that the additional compression of $Z_3$ does not seem to introduce additional error, and may even exhibit partial error cancellation. Even if the same were true in the case of $\tilde{Z}\tilde{D}Z$, the ``incorrect" compression subspace in which the orthogonal rotation is performed is the most likely cause of this discrepancy, and overwhelms any other considerations of the improved (due to the more appropriate choice of $U$) initial compression of $Z_3$.

\section{Conclusions}

In this work, we have examined five different approximation schemes for the perturbative triples correction in CCSD(T). Four of these five are entirely novel. For a subset of species taken from the W4-17 dataset, we have constructed a series of reaction energies using an extended version of the ANL$n$ reaction scheme. It was demonstrated that of the approximations studied here, including density-fitting of ERIs, rank-reduced compression of the $T_2$ amplitudes, Tucker-3 compression of the $Z_3/T_3$ intermediates in the (T) correction, and either Laplace transform or orthogonal rotation of the orbital eigenvalue denominators, only the $Z_3/T_3$ Tucker-3 decomposition was a limiting factor in the accuracy and cost of the resulting approximation schemes, and the number of projectors $U$ can be used to control the accuracy of the calculation. These five schemes, $\tilde{Z}T$, $\tilde{T}Z$ $\tilde{Z}\tilde{D}Z$, $\tilde{Z}^\prime\tilde{D}Z$, and $\tilde{Z}\tilde{D}\tilde{Z}$, are all capable of achieving sub-0.1 {\kJmol{}} mean absolute errors with modest numbers of projectors when compared against canonical, density-fitted DF-CCSD(T) calculations. In particular, $\tilde{Z}T$, novel to this work, demonstrates a favorable combination of accuracy and computational efficiency. Future work is needed to determine exactly how these components should be included in high-accuracy model chemistries, with particular emphasis on selecting the number of $Z_3/T_3$ projectors on a reaction-by-reaction basis, but the work here illustrates that these techniques have a place in some of our most accurate models of gas-phase thermochemistry and may yield order-of-magnitude improvements in computational cost of CCSD(T) with minimal and well-controlled error.

\begin{acknowledgments}
This work was supported by the US National Science Foundation under grant CHE-2143725. JHT acknowledges support from the SMU Moody School of Graduate and Advanced Studies. All calculations were performed on the ManeFrame III computing system at Southern Methodist University.
\end{acknowledgments}

\section*{Data Availability Statement}

The data that support the findings of this study are available within the article and its supplementary material. Available supplemental information files:
\begin{itemize}
\item \texttt{df\_error.xlsx}: absolute and relative errors of DF-CCSD(T) vs. CCSD(T) (aug-cc-pVTZ and aug-cc-pVQZ, frozen core).
\item \texttt{T2\_convergence\_error.xlsx} and \texttt{T2\_convergence\_raw\_data.xlsx}: rank-reduced $T_2$ amplitude convergence data.
\item \texttt{HOOI\_convergence\_error.xlsx} and \texttt{HOOI\_convergence\_raw\_data.xlsx}: HOOI convergence data.
\item \texttt{T3\_PVNZ\_FC\_raw\_data.xlsx}, \texttt{T3\_PCVNZ\_FC\_raw\_data.xlsx}, and \break \texttt{T3\_PCVNZ\_AE\_raw\_data.xlsx}: raw and extrapolated data for all reported results.
\item \texttt{supplemental.pdf}: factorized formulae for the implemented approximate CCSD(T) methods and additional supporting figures.
\end{itemize}

\bibliography{references}

\begin{thebibliography}{68}%
\makeatletter
\providecommand \@ifxundefined [1]{%
 \@ifx{#1\undefined}
}%
\providecommand \@ifnum [1]{%
 \ifnum #1\expandafter \@firstoftwo
 \else \expandafter \@secondoftwo
 \fi
}%
\providecommand \@ifx [1]{%
 \ifx #1\expandafter \@firstoftwo
 \else \expandafter \@secondoftwo
 \fi
}%
\providecommand \natexlab [1]{#1}%
\providecommand \enquote  [1]{``#1''}%
\providecommand \bibnamefont  [1]{#1}%
\providecommand \bibfnamefont [1]{#1}%
\providecommand \citenamefont [1]{#1}%
\providecommand \href@noop [0]{\@secondoftwo}%
\providecommand \href [0]{\begingroup \@sanitize@url \@href}%
\providecommand \@href[1]{\@@startlink{#1}\@@href}%
\providecommand \@@href[1]{\endgroup#1\@@endlink}%
\providecommand \@sanitize@url [0]{\catcode `\\12\catcode `\$12\catcode
  `\&12\catcode `\#12\catcode `\^12\catcode `\_12\catcode `\%12\relax}%
\providecommand \@@startlink[1]{}%
\providecommand \@@endlink[0]{}%
\providecommand \url  [0]{\begingroup\@sanitize@url \@url }%
\providecommand \@url [1]{\endgroup\@href {#1}{\urlprefix }}%
\providecommand \urlprefix  [0]{URL }%
\providecommand \Eprint [0]{\href }%
\providecommand \doibase [0]{https://doi.org/}%
\providecommand \selectlanguage [0]{\@gobble}%
\providecommand \bibinfo  [0]{\@secondoftwo}%
\providecommand \bibfield  [0]{\@secondoftwo}%
\providecommand \translation [1]{[#1]}%
\providecommand \BibitemOpen [0]{}%
\providecommand \bibitemStop [0]{}%
\providecommand \bibitemNoStop [0]{.\EOS\space}%
\providecommand \EOS [0]{\spacefactor3000\relax}%
\providecommand \BibitemShut  [1]{\csname bibitem#1\endcsname}%
\let\auto@bib@innerbib\@empty
\bibitem [{\citenamefont {Karton}\ \emph {et~al.}(2006)\citenamefont {Karton},
  \citenamefont {Rabinovich}, \citenamefont {Martin},\ and\ \citenamefont
  {Ruscic}}]{Karton2006}%
  \BibitemOpen
  \bibfield  {author} {\bibinfo {author} {\bibfnamefont {A.}~\bibnamefont
  {Karton}}, \bibinfo {author} {\bibfnamefont {E.}~\bibnamefont {Rabinovich}},
  \bibinfo {author} {\bibfnamefont {J.~M.~L.}\ \bibnamefont {Martin}},\ and\
  \bibinfo {author} {\bibfnamefont {B.}~\bibnamefont {Ruscic}},\ }\bibfield
  {title} {\enquote {\bibinfo {title} {W4 theory for computational
  thermochemistry: In pursuit of confident sub-{kJ}/mol predictions},}\ }\href
  {https://doi.org/10.1063/1.2348881} {\bibfield  {journal} {\bibinfo
  {journal} {The Journal of Chemical Physics}\ }\textbf {\bibinfo {volume}
  {125}} (\bibinfo {year} {2006}),\ 10.1063/1.2348881}\BibitemShut {NoStop}%
\bibitem [{\citenamefont {Feller}, \citenamefont {Peterson},\ and\
  \citenamefont {Dixon}(2008)}]{Feller2008}%
  \BibitemOpen
  \bibfield  {author} {\bibinfo {author} {\bibfnamefont {D.}~\bibnamefont
  {Feller}}, \bibinfo {author} {\bibfnamefont {K.~A.}\ \bibnamefont
  {Peterson}},\ and\ \bibinfo {author} {\bibfnamefont {D.~A.}\ \bibnamefont
  {Dixon}},\ }\bibfield  {title} {\enquote {\bibinfo {title} {A survey of
  factors contributing to accurate theoretical predictions of atomization
  energies and molecular structures},}\ }\href
  {https://doi.org/10.1063/1.3008061} {\bibfield  {journal} {\bibinfo
  {journal} {The Journal of Chemical Physics}\ }\textbf {\bibinfo {volume}
  {129}} (\bibinfo {year} {2008}),\ 10.1063/1.3008061}\BibitemShut {NoStop}%
\bibitem [{\citenamefont {Harding}\ \emph {et~al.}(2008)\citenamefont
  {Harding}, \citenamefont {V{\'{a}}zquez}, \citenamefont {Ruscic},
  \citenamefont {Wilson}, \citenamefont {Gauss},\ and\ \citenamefont
  {Stanton}}]{Harding2008}%
  \BibitemOpen
  \bibfield  {author} {\bibinfo {author} {\bibfnamefont {M.~E.}\ \bibnamefont
  {Harding}}, \bibinfo {author} {\bibfnamefont {J.}~\bibnamefont
  {V{\'{a}}zquez}}, \bibinfo {author} {\bibfnamefont {B.}~\bibnamefont
  {Ruscic}}, \bibinfo {author} {\bibfnamefont {A.~K.}\ \bibnamefont {Wilson}},
  \bibinfo {author} {\bibfnamefont {J.}~\bibnamefont {Gauss}},\ and\ \bibinfo
  {author} {\bibfnamefont {J.~F.}\ \bibnamefont {Stanton}},\ }\bibfield
  {title} {\enquote {\bibinfo {title} {High-accuracy extrapolated \emph{ab
  initio} thermochemistry. {III}. additional improvements and overview},}\
  }\href {https://doi.org/10.1063/1.2835612} {\bibfield  {journal} {\bibinfo
  {journal} {The Journal of Chemical Physics}\ }\textbf {\bibinfo {volume}
  {128}} (\bibinfo {year} {2008}),\ 10.1063/1.2835612}\BibitemShut {NoStop}%
\bibitem [{\citenamefont {Boese}\ \emph {et~al.}(2004)\citenamefont {Boese},
  \citenamefont {Oren}, \citenamefont {Atasoylu}, \citenamefont {Martin},
  \citenamefont {K\'{a}llay},\ and\ \citenamefont {Gauss}}]{Boese-JCP-2004}%
  \BibitemOpen
  \bibfield  {author} {\bibinfo {author} {\bibfnamefont {A.~D.}\ \bibnamefont
  {Boese}}, \bibinfo {author} {\bibfnamefont {M.}~\bibnamefont {Oren}},
  \bibinfo {author} {\bibfnamefont {O.}~\bibnamefont {Atasoylu}}, \bibinfo
  {author} {\bibfnamefont {J.~M.~L.}\ \bibnamefont {Martin}}, \bibinfo {author}
  {\bibfnamefont {M.}~\bibnamefont {K\'{a}llay}},\ and\ \bibinfo {author}
  {\bibfnamefont {J.}~\bibnamefont {Gauss}},\ }\bibfield  {title} {\enquote
  {\bibinfo {title} {{W}3 theory: Robust computational thermochemistry in the
  k{J}/mol accuracy range},}\ }\href {https://doi.org/10.1063/1.1638736}
  {\bibfield  {journal} {\bibinfo  {journal} {The Journal of Chemical Physics}\
  }\textbf {\bibinfo {volume} {120}},\ \bibinfo {pages} {4129--4141} (\bibinfo
  {year} {2004})}\BibitemShut {NoStop}%
\bibitem [{\citenamefont {Sylvetsky}\ \emph {et~al.}(2016)\citenamefont
  {Sylvetsky}, \citenamefont {Peterson}, \citenamefont {Karton},\ and\
  \citenamefont {Martin}}]{Sylvetsky2016}%
  \BibitemOpen
  \bibfield  {author} {\bibinfo {author} {\bibfnamefont {N.}~\bibnamefont
  {Sylvetsky}}, \bibinfo {author} {\bibfnamefont {K.~A.}\ \bibnamefont
  {Peterson}}, \bibinfo {author} {\bibfnamefont {A.}~\bibnamefont {Karton}},\
  and\ \bibinfo {author} {\bibfnamefont {J.~M.~L.}\ \bibnamefont {Martin}},\
  }\bibfield  {title} {\enquote {\bibinfo {title} {Toward a w4-f12 approach:
  Can explicitly correlated and orbital-based \emph{ab initio} {CCSD}(t) limits
  be reconciled?}}\ }\href {https://doi.org/10.1063/1.4952410} {\bibfield
  {journal} {\bibinfo  {journal} {The Journal of Chemical Physics}\ }\textbf
  {\bibinfo {volume} {144}} (\bibinfo {year} {2016}),\
  10.1063/1.4952410}\BibitemShut {NoStop}%
\bibitem [{\citenamefont {Tajti}\ \emph {et~al.}(2004)\citenamefont {Tajti},
  \citenamefont {Szalay}, \citenamefont {Cs{\'{a}}sz{\'{a}}r}, \citenamefont
  {K{\'{a}}llay}, \citenamefont {Gauss}, \citenamefont {Valeev}, \citenamefont
  {Flowers}, \citenamefont {V{\'{a}}zquez},\ and\ \citenamefont
  {Stanton}}]{Tajti2004}%
  \BibitemOpen
  \bibfield  {author} {\bibinfo {author} {\bibfnamefont {A.}~\bibnamefont
  {Tajti}}, \bibinfo {author} {\bibfnamefont {P.~G.}\ \bibnamefont {Szalay}},
  \bibinfo {author} {\bibfnamefont {A.~G.}\ \bibnamefont
  {Cs{\'{a}}sz{\'{a}}r}}, \bibinfo {author} {\bibfnamefont {M.}~\bibnamefont
  {K{\'{a}}llay}}, \bibinfo {author} {\bibfnamefont {J.}~\bibnamefont {Gauss}},
  \bibinfo {author} {\bibfnamefont {E.~F.}\ \bibnamefont {Valeev}}, \bibinfo
  {author} {\bibfnamefont {B.~A.}\ \bibnamefont {Flowers}}, \bibinfo {author}
  {\bibfnamefont {J.}~\bibnamefont {V{\'{a}}zquez}},\ and\ \bibinfo {author}
  {\bibfnamefont {J.~F.}\ \bibnamefont {Stanton}},\ }\bibfield  {title}
  {\enquote {\bibinfo {title} {{HEAT}: High accuracy extrapolated \emph{ab
  initio} thermochemistry},}\ }\href {https://doi.org/10.1063/1.1811608}
  {\bibfield  {journal} {\bibinfo  {journal} {The Journal of Chemical Physics}\
  }\textbf {\bibinfo {volume} {121}},\ \bibinfo {pages} {11599--11613}
  (\bibinfo {year} {2004})}\BibitemShut {NoStop}%
\bibitem [{\citenamefont {Bomble}\ \emph {et~al.}(2006)\citenamefont {Bomble},
  \citenamefont {V{\'{a}}zquez}, \citenamefont {K{\'{a}}llay}, \citenamefont
  {Michauk}, \citenamefont {Szalay}, \citenamefont {Cs{\'{a}}sz{\'{a}}r},
  \citenamefont {Gauss},\ and\ \citenamefont {Stanton}}]{Bomble2006}%
  \BibitemOpen
  \bibfield  {author} {\bibinfo {author} {\bibfnamefont {Y.~J.}\ \bibnamefont
  {Bomble}}, \bibinfo {author} {\bibfnamefont {J.}~\bibnamefont
  {V{\'{a}}zquez}}, \bibinfo {author} {\bibfnamefont {M.}~\bibnamefont
  {K{\'{a}}llay}}, \bibinfo {author} {\bibfnamefont {C.}~\bibnamefont
  {Michauk}}, \bibinfo {author} {\bibfnamefont {P.~G.}\ \bibnamefont {Szalay}},
  \bibinfo {author} {\bibfnamefont {A.~G.}\ \bibnamefont
  {Cs{\'{a}}sz{\'{a}}r}}, \bibinfo {author} {\bibfnamefont {J.}~\bibnamefont
  {Gauss}},\ and\ \bibinfo {author} {\bibfnamefont {J.~F.}\ \bibnamefont
  {Stanton}},\ }\bibfield  {title} {\enquote {\bibinfo {title} {High-accuracy
  extrapolated \emph{ab initio} thermochemistry. {II}. minor improvements to
  the protocol and a vital simplification},}\ }\href
  {https://doi.org/10.1063/1.2206789} {\bibfield  {journal} {\bibinfo
  {journal} {The Journal of Chemical Physics}\ }\textbf {\bibinfo {volume}
  {125}} (\bibinfo {year} {2006}),\ 10.1063/1.2206789}\BibitemShut {NoStop}%
\bibitem [{\citenamefont {Thorpe}\ \emph {et~al.}(2019)\citenamefont {Thorpe},
  \citenamefont {Lopez}, \citenamefont {Nguyen}, \citenamefont {Baraban},
  \citenamefont {Bross}, \citenamefont {Ruscic},\ and\ \citenamefont
  {Stanton}}]{Thorpe2019}%
  \BibitemOpen
  \bibfield  {author} {\bibinfo {author} {\bibfnamefont {J.~H.}\ \bibnamefont
  {Thorpe}}, \bibinfo {author} {\bibfnamefont {C.~A.}\ \bibnamefont {Lopez}},
  \bibinfo {author} {\bibfnamefont {T.~L.}\ \bibnamefont {Nguyen}}, \bibinfo
  {author} {\bibfnamefont {J.~H.}\ \bibnamefont {Baraban}}, \bibinfo {author}
  {\bibfnamefont {D.~H.}\ \bibnamefont {Bross}}, \bibinfo {author}
  {\bibfnamefont {B.}~\bibnamefont {Ruscic}},\ and\ \bibinfo {author}
  {\bibfnamefont {J.~F.}\ \bibnamefont {Stanton}},\ }\bibfield  {title}
  {\enquote {\bibinfo {title} {High-accuracy extrapolated \emph{ab initio}
  thermochemistry. {IV}. a modified recipe for computational efficiency},}\
  }\href {https://doi.org/10.1063/1.5095937} {\bibfield  {journal} {\bibinfo
  {journal} {The Journal of Chemical Physics}\ }\textbf {\bibinfo {volume}
  {150}} (\bibinfo {year} {2019}),\ 10.1063/1.5095937}\BibitemShut {NoStop}%
\bibitem [{\citenamefont {Thorpe}\ \emph {et~al.}(2021)\citenamefont {Thorpe},
  \citenamefont {Kilburn}, \citenamefont {Feller}, \citenamefont {Changala},
  \citenamefont {Bross}, \citenamefont {Ruscic},\ and\ \citenamefont
  {Stanton}}]{Thorpe-JCP-2021}%
  \BibitemOpen
  \bibfield  {author} {\bibinfo {author} {\bibfnamefont {J.~H.}\ \bibnamefont
  {Thorpe}}, \bibinfo {author} {\bibfnamefont {J.~L.}\ \bibnamefont {Kilburn}},
  \bibinfo {author} {\bibfnamefont {D.}~\bibnamefont {Feller}}, \bibinfo
  {author} {\bibfnamefont {P.~B.}\ \bibnamefont {Changala}}, \bibinfo {author}
  {\bibfnamefont {D.~H.}\ \bibnamefont {Bross}}, \bibinfo {author}
  {\bibfnamefont {B.}~\bibnamefont {Ruscic}},\ and\ \bibinfo {author}
  {\bibfnamefont {J.~F.}\ \bibnamefont {Stanton}},\ }\bibfield  {title}
  {\enquote {\bibinfo {title} {Elaborated thermochemical treatment of hf, co,
  n2, and h2o: Insight into heat and its extensions},}\ }\href
  {https://doi.org/10.1063/5.0069322} {\bibfield  {journal} {\bibinfo
  {journal} {The Journal of Chemical Physics}\ }\textbf {\bibinfo {volume}
  {155}},\ \bibinfo {pages} {184109} (\bibinfo {year} {2021})}\BibitemShut
  {NoStop}%
\bibitem [{\citenamefont {Klippenstein}, \citenamefont {Harding},\ and\
  \citenamefont {Ruscic}(2017)}]{Klippenstein2017}%
  \BibitemOpen
  \bibfield  {author} {\bibinfo {author} {\bibfnamefont {S.~J.}\ \bibnamefont
  {Klippenstein}}, \bibinfo {author} {\bibfnamefont {L.~B.}\ \bibnamefont
  {Harding}},\ and\ \bibinfo {author} {\bibfnamefont {B.}~\bibnamefont
  {Ruscic}},\ }\bibfield  {title} {\enquote {\bibinfo {title} {Ab initio
  computations and active thermochemical tables hand in hand: Heats of
  formation of core combustion species},}\ }\href
  {https://doi.org/10.1021/acs.jpca.7b05945} {\bibfield  {journal} {\bibinfo
  {journal} {The Journal of Physical Chemistry A}\ }\textbf {\bibinfo {volume}
  {121}},\ \bibinfo {pages} {6580–6602} (\bibinfo {year} {2017})}\BibitemShut
  {NoStop}%
\bibitem [{\citenamefont {Ruscic}(2014)}]{Ruscic-IJQC-2014}%
  \BibitemOpen
  \bibfield  {author} {\bibinfo {author} {\bibfnamefont {B.}~\bibnamefont
  {Ruscic}},\ }\bibfield  {title} {\enquote {\bibinfo {title} {Uncertainty
  quantification in thermochemistry, benchmarking electronic structure
  computations, and active thermochemical tables},}\ }\href
  {https://doi.org/10.1002/qua.24605} {\bibfield  {journal} {\bibinfo
  {journal} {International Journal of Quantum Chemistry}\ }\textbf {\bibinfo
  {volume} {114}},\ \bibinfo {pages} {1097--1101} (\bibinfo {year}
  {2014})}\BibitemShut {NoStop}%
\bibitem [{\citenamefont {East}\ and\ \citenamefont {Allen}(1993)}]{East1993}%
  \BibitemOpen
  \bibfield  {author} {\bibinfo {author} {\bibfnamefont {A.~L.~L.}\
  \bibnamefont {East}}\ and\ \bibinfo {author} {\bibfnamefont {W.~D.}\
  \bibnamefont {Allen}},\ }\bibfield  {title} {\enquote {\bibinfo {title} {The
  heat of formation of {NCO}},}\ }\href {https://doi.org/10.1063/1.466062}
  {\bibfield  {journal} {\bibinfo  {journal} {The Journal of Chemical Physics}\
  }\textbf {\bibinfo {volume} {99}},\ \bibinfo {pages} {4638--4650} (\bibinfo
  {year} {1993})}\BibitemShut {NoStop}%
\bibitem [{\citenamefont {Cs\'{a}sz\'{a}r}, \citenamefont {Allen},\ and\
  \citenamefont {Schaefer}(1998)}]{Csaszar-JCP-1998}%
  \BibitemOpen
  \bibfield  {author} {\bibinfo {author} {\bibfnamefont {A.~G.}\ \bibnamefont
  {Cs\'{a}sz\'{a}r}}, \bibinfo {author} {\bibfnamefont {W.~D.}\ \bibnamefont
  {Allen}},\ and\ \bibinfo {author} {\bibfnamefont {H.~F.}\ \bibnamefont
  {Schaefer}},\ }\bibfield  {title} {\enquote {\bibinfo {title} {In pursuit of
  the {\it ab initio} limit for conformational energy prototypes},}\ }\href
  {https://doi.org/10.1063/1.476449} {\bibfield  {journal} {\bibinfo  {journal}
  {The Journal of Chemical Physics}\ }\textbf {\bibinfo {volume} {108}},\
  \bibinfo {pages} {9751--9764} (\bibinfo {year} {1998})}\BibitemShut {NoStop}%
\bibitem [{\citenamefont {Kenny}, \citenamefont {Allen},\ and\ \citenamefont
  {Schaefer}(2003)}]{Kenny-JCP-2003}%
  \BibitemOpen
  \bibfield  {author} {\bibinfo {author} {\bibfnamefont {J.~P.}\ \bibnamefont
  {Kenny}}, \bibinfo {author} {\bibfnamefont {W.~D.}\ \bibnamefont {Allen}},\
  and\ \bibinfo {author} {\bibfnamefont {H.~F.}\ \bibnamefont {Schaefer}},\
  }\bibfield  {title} {\enquote {\bibinfo {title} {Complete basis set limit
  studies of conventional and {R12} correlation methods: The silicon dicarbide
  ({SiC$-2$}) barrier to linearity},}\ }\href
  {https://doi.org/10.1063/1.1558533} {\bibfield  {journal} {\bibinfo
  {journal} {The Journal of Chemical Physics}\ }\textbf {\bibinfo {volume}
  {118}},\ \bibinfo {pages} {7353--7365} (\bibinfo {year} {2003})}\BibitemShut
  {NoStop}%
\bibitem [{\citenamefont {Schuurman}\ \emph {et~al.}(2004)\citenamefont
  {Schuurman}, \citenamefont {Muir}, \citenamefont {Allen},\ and\ \citenamefont
  {Schaefer}}]{Schuurman-JCP-2004}%
  \BibitemOpen
  \bibfield  {author} {\bibinfo {author} {\bibfnamefont {M.~S.}\ \bibnamefont
  {Schuurman}}, \bibinfo {author} {\bibfnamefont {S.~R.}\ \bibnamefont {Muir}},
  \bibinfo {author} {\bibfnamefont {W.~D.}\ \bibnamefont {Allen}},\ and\
  \bibinfo {author} {\bibfnamefont {H.~F.}\ \bibnamefont {Schaefer}},\
  }\bibfield  {title} {\enquote {\bibinfo {title} {Toward subchemical accuracy
  in computational thermochemistry: Focal point analysis of the heat of
  formation of {NCO} and [{H},{N},{C},{O}] isomers},}\ }\href
  {https://doi.org/10.1063/1.1707013} {\bibfield  {journal} {\bibinfo
  {journal} {The Journal of Chemical Physics}\ }\textbf {\bibinfo {volume}
  {120}},\ \bibinfo {pages} {11586--11599} (\bibinfo {year}
  {2004})}\BibitemShut {NoStop}%
\bibitem [{\citenamefont {Feller}, \citenamefont {Peterson},\ and\
  \citenamefont {Dixon}(2012)}]{Feller2012}%
  \BibitemOpen
  \bibfield  {author} {\bibinfo {author} {\bibfnamefont {D.}~\bibnamefont
  {Feller}}, \bibinfo {author} {\bibfnamefont {K.~A.}\ \bibnamefont
  {Peterson}},\ and\ \bibinfo {author} {\bibfnamefont {D.~A.}\ \bibnamefont
  {Dixon}},\ }\bibfield  {title} {\enquote {\bibinfo {title} {Further
  benchmarks of a composite, convergent, statistically calibrated
  coupled-cluster-based approach for thermochemical and spectroscopic
  studies},}\ }\href {https://doi.org/10.1080/00268976.2012.684897} {\bibfield
  {journal} {\bibinfo  {journal} {Molecular Physics}\ }\textbf {\bibinfo
  {volume} {110}},\ \bibinfo {pages} {2381--2399} (\bibinfo {year}
  {2012})}\BibitemShut {NoStop}%
\bibitem [{\citenamefont {Peterson}, \citenamefont {Feller},\ and\
  \citenamefont {Dixon}(2012)}]{Peterson2012}%
  \BibitemOpen
  \bibfield  {author} {\bibinfo {author} {\bibfnamefont {K.~A.}\ \bibnamefont
  {Peterson}}, \bibinfo {author} {\bibfnamefont {D.}~\bibnamefont {Feller}},\
  and\ \bibinfo {author} {\bibfnamefont {D.~A.}\ \bibnamefont {Dixon}},\
  }\bibfield  {title} {\enquote {\bibinfo {title} {Chemical accuracy in ab
  initio thermochemistry and spectroscopy: current strategies and future
  challenges},}\ }\href {https://doi.org/10.1007/s00214-011-1079-5} {\bibfield
  {journal} {\bibinfo  {journal} {Theoretical Chemistry Accounts}\ }\textbf
  {\bibinfo {volume} {131}} (\bibinfo {year} {2012}),\
  10.1007/s00214-011-1079-5}\BibitemShut {NoStop}%
\bibitem [{\citenamefont {Dixon}, \citenamefont {Feller},\ and\ \citenamefont
  {Peterson}(2012)}]{Dixon2012}%
  \BibitemOpen
  \bibfield  {author} {\bibinfo {author} {\bibfnamefont {D.~A.}\ \bibnamefont
  {Dixon}}, \bibinfo {author} {\bibfnamefont {D.}~\bibnamefont {Feller}},\ and\
  \bibinfo {author} {\bibfnamefont {K.~A.}\ \bibnamefont {Peterson}},\
  }\bibfield  {title} {\enquote {\bibinfo {title} {A practical guide to
  reliable first principles computational thermochemistry predictions across
  the periodic table},}\ }in\ \href
  {https://doi.org/10.1016/b978-0-444-59440-2.00001-6} {\emph {\bibinfo
  {booktitle} {Annual Reports in Computational Chemistry Volume 8}}}\ (\bibinfo
   {publisher} {Elsevier},\ \bibinfo {year} {2012})\ pp.\ \bibinfo {pages}
  {1--28}\BibitemShut {NoStop}%
\bibitem [{\citenamefont {Feller}, \citenamefont {Peterson},\ and\
  \citenamefont {Ruscic}(2013)}]{Feller-TCA-2013}%
  \BibitemOpen
  \bibfield  {author} {\bibinfo {author} {\bibfnamefont {D.}~\bibnamefont
  {Feller}}, \bibinfo {author} {\bibfnamefont {K.~A.}\ \bibnamefont
  {Peterson}},\ and\ \bibinfo {author} {\bibfnamefont {B.}~\bibnamefont
  {Ruscic}},\ }\bibfield  {title} {\enquote {\bibinfo {title} {Improved
  accuracy benchmarks of small molecules using correlation consistent basis
  sets},}\ }\href {https://doi.org/10.1007/s00214-013-1407-z} {\bibfield
  {journal} {\bibinfo  {journal} {Theoretical Chemistry Accounts}\ }\textbf
  {\bibinfo {volume} {133}} (\bibinfo {year} {2013}),\
  10.1007/s00214-013-1407-z}\BibitemShut {NoStop}%
\bibitem [{\citenamefont {Raghavachari}\ \emph {et~al.}(1989)\citenamefont
  {Raghavachari}, \citenamefont {Trucks}, \citenamefont {Pople},\ and\
  \citenamefont {Head-Gordon}}]{T_Raghavachari1989}%
  \BibitemOpen
  \bibfield  {author} {\bibinfo {author} {\bibfnamefont {K.}~\bibnamefont
  {Raghavachari}}, \bibinfo {author} {\bibfnamefont {G.~W.}\ \bibnamefont
  {Trucks}}, \bibinfo {author} {\bibfnamefont {J.~A.}\ \bibnamefont {Pople}},\
  and\ \bibinfo {author} {\bibfnamefont {M.}~\bibnamefont {Head-Gordon}},\
  }\bibfield  {title} {\enquote {\bibinfo {title} {A fifth-order perturbation
  comparison of electron correlation theories},}\ }\href
  {https://doi.org/10.1016/s0009-2614(89)87395-6} {\bibfield  {journal}
  {\bibinfo  {journal} {Chemical Physics Letters}\ }\textbf {\bibinfo {volume}
  {157}},\ \bibinfo {pages} {479--483} (\bibinfo {year} {1989})}\BibitemShut
  {NoStop}%
\bibitem [{\citenamefont {Purvis}\ and\ \citenamefont
  {Bartlett}(1982)}]{Purvis-JCP-1982}%
  \BibitemOpen
  \bibfield  {author} {\bibinfo {author} {\bibfnamefont {G.~D.}\ \bibnamefont
  {Purvis}}\ and\ \bibinfo {author} {\bibfnamefont {R.~J.}\ \bibnamefont
  {Bartlett}},\ }\bibfield  {title} {\enquote {\bibinfo {title} {A full
  coupled‐cluster singles and doubles model: The inclusion of disconnected
  triples},}\ }\href {https://doi.org/10.1063/1.443164} {\bibfield  {journal}
  {\bibinfo  {journal} {The Journal of Chemical Physics}\ }\textbf {\bibinfo
  {volume} {76}},\ \bibinfo {pages} {1910--1918} (\bibinfo {year}
  {1982})}\BibitemShut {NoStop}%
\bibitem [{\citenamefont {Karton}\ \emph {et~al.}(2013)\citenamefont {Karton},
  \citenamefont {Chan}, \citenamefont {Raghavachari},\ and\ \citenamefont
  {Radom}}]{Karton2013}%
  \BibitemOpen
  \bibfield  {author} {\bibinfo {author} {\bibfnamefont {A.}~\bibnamefont
  {Karton}}, \bibinfo {author} {\bibfnamefont {B.}~\bibnamefont {Chan}},
  \bibinfo {author} {\bibfnamefont {K.}~\bibnamefont {Raghavachari}},\ and\
  \bibinfo {author} {\bibfnamefont {L.}~\bibnamefont {Radom}},\ }\bibfield
  {title} {\enquote {\bibinfo {title} {Evaluation of the heats of formation of
  corannulene and $\text{C}_{60}$ by means of high-level theoretical
  procedures},}\ }\href {https://doi.org/10.1021/jp312585r} {\bibfield
  {journal} {\bibinfo  {journal} {The Journal of Physical Chemistry A}\
  }\textbf {\bibinfo {volume} {117}},\ \bibinfo {pages} {1834--1842} (\bibinfo
  {year} {2013})}\BibitemShut {NoStop}%
\bibitem [{\citenamefont {Karton}, \citenamefont {Schreiner},\ and\
  \citenamefont {Martin}(2015)}]{Karton2015}%
  \BibitemOpen
  \bibfield  {author} {\bibinfo {author} {\bibfnamefont {A.}~\bibnamefont
  {Karton}}, \bibinfo {author} {\bibfnamefont {P.~R.}\ \bibnamefont
  {Schreiner}},\ and\ \bibinfo {author} {\bibfnamefont {J.~M.~L.}\ \bibnamefont
  {Martin}},\ }\bibfield  {title} {\enquote {\bibinfo {title} {Heats of
  formation of platonic hydrocarbon cages by means of high-level thermochemical
  procedures},}\ }\href {https://doi.org/10.1002/jcc.23963} {\bibfield
  {journal} {\bibinfo  {journal} {Journal of Computational Chemistry}\ }\textbf
  {\bibinfo {volume} {37}},\ \bibinfo {pages} {49--58} (\bibinfo {year}
  {2015})}\BibitemShut {NoStop}%
\bibitem [{\citenamefont {Manna}\ and\ \citenamefont
  {Martin}(2015)}]{Manna2015}%
  \BibitemOpen
  \bibfield  {author} {\bibinfo {author} {\bibfnamefont {D.}~\bibnamefont
  {Manna}}\ and\ \bibinfo {author} {\bibfnamefont {J.~M.~L.}\ \bibnamefont
  {Martin}},\ }\bibfield  {title} {\enquote {\bibinfo {title} {What are the
  ground state structures of $\text{C}_{20}$ and $\text{C}_{24}$? an explicitly
  correlated ab initio approach},}\ }\href
  {https://doi.org/10.1021/acs.jpca.5b10266} {\bibfield  {journal} {\bibinfo
  {journal} {The Journal of Physical Chemistry A}\ }\textbf {\bibinfo {volume}
  {120}},\ \bibinfo {pages} {153--160} (\bibinfo {year} {2015})}\BibitemShut
  {NoStop}%
\bibitem [{\citenamefont {Riplinger}\ and\ \citenamefont
  {Neese}(2013)}]{Riplinger2013_jan}%
  \BibitemOpen
  \bibfield  {author} {\bibinfo {author} {\bibfnamefont {C.}~\bibnamefont
  {Riplinger}}\ and\ \bibinfo {author} {\bibfnamefont {F.}~\bibnamefont
  {Neese}},\ }\bibfield  {title} {\enquote {\bibinfo {title} {An efficient and
  near linear scaling pair natural orbital based local coupled cluster
  method},}\ }\href {https://doi.org/10.1063/1.4773581} {\bibfield  {journal}
  {\bibinfo  {journal} {The Journal of Chemical Physics}\ }\textbf {\bibinfo
  {volume} {138}} (\bibinfo {year} {2013}),\ 10.1063/1.4773581}\BibitemShut
  {NoStop}%
\bibitem [{\citenamefont {Riplinger}\ \emph {et~al.}(2013)\citenamefont
  {Riplinger}, \citenamefont {Sandhoefer}, \citenamefont {Hansen},\ and\
  \citenamefont {Neese}}]{Riplinger2013_oct}%
  \BibitemOpen
  \bibfield  {author} {\bibinfo {author} {\bibfnamefont {C.}~\bibnamefont
  {Riplinger}}, \bibinfo {author} {\bibfnamefont {B.}~\bibnamefont
  {Sandhoefer}}, \bibinfo {author} {\bibfnamefont {A.}~\bibnamefont {Hansen}},\
  and\ \bibinfo {author} {\bibfnamefont {F.}~\bibnamefont {Neese}},\ }\bibfield
   {title} {\enquote {\bibinfo {title} {Natural triple excitations in local
  coupled cluster calculations with pair natural orbitals},}\ }\href
  {https://doi.org/10.1063/1.4821834} {\bibfield  {journal} {\bibinfo
  {journal} {The Journal of Chemical Physics}\ }\textbf {\bibinfo {volume}
  {139}} (\bibinfo {year} {2013}),\ 10.1063/1.4821834}\BibitemShut {NoStop}%
\bibitem [{\citenamefont {Liakos}\ \emph {et~al.}(2015)\citenamefont {Liakos},
  \citenamefont {Sparta}, \citenamefont {Kesharwani}, \citenamefont {Martin},\
  and\ \citenamefont {Neese}}]{Liakos-JCTC-2015}%
  \BibitemOpen
  \bibfield  {author} {\bibinfo {author} {\bibfnamefont {D.~G.}\ \bibnamefont
  {Liakos}}, \bibinfo {author} {\bibfnamefont {M.}~\bibnamefont {Sparta}},
  \bibinfo {author} {\bibfnamefont {M.~K.}\ \bibnamefont {Kesharwani}},
  \bibinfo {author} {\bibfnamefont {J.~M.~L.}\ \bibnamefont {Martin}},\ and\
  \bibinfo {author} {\bibfnamefont {F.}~\bibnamefont {Neese}},\ }\bibfield
  {title} {\enquote {\bibinfo {title} {Exploring the accuracy limits of local
  pair natural orbital coupled-cluster theory},}\ }\href
  {https://doi.org/10.1021/ct501129s} {\bibfield  {journal} {\bibinfo
  {journal} {Journal of Chemical Theory and Computation}\ }\textbf {\bibinfo
  {volume} {11}},\ \bibinfo {pages} {1525--1539} (\bibinfo {year}
  {2015})}\BibitemShut {NoStop}%
\bibitem [{\citenamefont {Ma}\ and\ \citenamefont {Werner}(2018)}]{Ma2018}%
  \BibitemOpen
  \bibfield  {author} {\bibinfo {author} {\bibfnamefont {Q.}~\bibnamefont
  {Ma}}\ and\ \bibinfo {author} {\bibfnamefont {H.-J.}\ \bibnamefont
  {Werner}},\ }\bibfield  {title} {\enquote {\bibinfo {title} {Explicitly
  correlated local coupled-cluster methods using pair natural orbitals},}\
  }\href {https://doi.org/10.1002/wcms.1371} {\bibfield  {journal} {\bibinfo
  {journal} {{WIREs} Computational Molecular Science}\ }\textbf {\bibinfo
  {volume} {8}} (\bibinfo {year} {2018}),\ 10.1002/wcms.1371}\BibitemShut
  {NoStop}%
\bibitem [{\citenamefont {Nagy}\ and\ \citenamefont
  {K{\'{a}}llay}(2019)}]{Nagy2019}%
  \BibitemOpen
  \bibfield  {author} {\bibinfo {author} {\bibfnamefont {P.~R.}\ \bibnamefont
  {Nagy}}\ and\ \bibinfo {author} {\bibfnamefont {M.}~\bibnamefont
  {K{\'{a}}llay}},\ }\bibfield  {title} {\enquote {\bibinfo {title}
  {Approaching the basis set limit of {CCSD}(t) energies for large molecules
  with local natural orbital coupled-cluster methods},}\ }\href
  {https://doi.org/10.1021/acs.jctc.9b00511} {\bibfield  {journal} {\bibinfo
  {journal} {Journal of Chemical Theory and Computation}\ }\textbf {\bibinfo
  {volume} {15}},\ \bibinfo {pages} {5275--5298} (\bibinfo {year}
  {2019})}\BibitemShut {NoStop}%
\bibitem [{\citenamefont {Karton}, \citenamefont {Taylor},\ and\ \citenamefont
  {Martin}(2007)}]{Karton2007}%
  \BibitemOpen
  \bibfield  {author} {\bibinfo {author} {\bibfnamefont {A.}~\bibnamefont
  {Karton}}, \bibinfo {author} {\bibfnamefont {P.~R.}\ \bibnamefont {Taylor}},\
  and\ \bibinfo {author} {\bibfnamefont {J.~M.~L.}\ \bibnamefont {Martin}},\
  }\bibfield  {title} {\enquote {\bibinfo {title} {Basis set convergence of
  post-{CCSD} contributions to molecular atomization energies},}\ }\href
  {https://doi.org/10.1063/1.2755751} {\bibfield  {journal} {\bibinfo
  {journal} {The Journal of Chemical Physics}\ }\textbf {\bibinfo {volume}
  {127}} (\bibinfo {year} {2007}),\ 10.1063/1.2755751}\BibitemShut {NoStop}%
\bibitem [{\citenamefont {Feller}, \citenamefont {Peterson},\ and\
  \citenamefont {Hill}(2011)}]{Feller2011}%
  \BibitemOpen
  \bibfield  {author} {\bibinfo {author} {\bibfnamefont {D.}~\bibnamefont
  {Feller}}, \bibinfo {author} {\bibfnamefont {K.~A.}\ \bibnamefont
  {Peterson}},\ and\ \bibinfo {author} {\bibfnamefont {J.~G.}\ \bibnamefont
  {Hill}},\ }\bibfield  {title} {\enquote {\bibinfo {title} {On the
  effectiveness of {CCSD}(t) complete basis set extrapolations for atomization
  energies},}\ }\href {https://doi.org/10.1063/1.3613639} {\bibfield  {journal}
  {\bibinfo  {journal} {The Journal of Chemical Physics}\ }\textbf {\bibinfo
  {volume} {135}} (\bibinfo {year} {2011}),\ 10.1063/1.3613639}\BibitemShut
  {NoStop}%
\bibitem [{\citenamefont {Feller}(2013)}]{Feller2013}%
  \BibitemOpen
  \bibfield  {author} {\bibinfo {author} {\bibfnamefont {D.}~\bibnamefont
  {Feller}},\ }\bibfield  {title} {\enquote {\bibinfo {title} {Benchmarks of
  improved complete basis set extrapolation schemes designed for standard
  {CCSD}(t) atomization energies},}\ }\href {https://doi.org/10.1063/1.4791560}
  {\bibfield  {journal} {\bibinfo  {journal} {The Journal of Chemical Physics}\
  }\textbf {\bibinfo {volume} {138}} (\bibinfo {year} {2013}),\
  10.1063/1.4791560}\BibitemShut {NoStop}%
\bibitem [{\citenamefont {Boys}\ and\ \citenamefont
  {Shavitt}(1959)}]{boys1959}%
  \BibitemOpen
  \bibfield  {author} {\bibinfo {author} {\bibfnamefont {S.~F.}\ \bibnamefont
  {Boys}}\ and\ \bibinfo {author} {\bibfnamefont {I.}~\bibnamefont {Shavitt}},\
  }\href@noop {} {\bibfield  {journal} {\bibinfo  {journal} {University of
  Wisconsin Naval Research Laboratory Report}\ } (\bibinfo {year}
  {1959})}\BibitemShut {NoStop}%
\bibitem [{\citenamefont {Vahtras}, \citenamefont {Alml\"{o}f},\ and\
  \citenamefont {Feyereisen}(1993)}]{Vahtras1993}%
  \BibitemOpen
  \bibfield  {author} {\bibinfo {author} {\bibfnamefont {O.}~\bibnamefont
  {Vahtras}}, \bibinfo {author} {\bibfnamefont {J.}~\bibnamefont
  {Alml\"{o}f}},\ and\ \bibinfo {author} {\bibfnamefont {M.}~\bibnamefont
  {Feyereisen}},\ }\bibfield  {title} {\enquote {\bibinfo {title} {Integral
  approximations for lcao-scf calculations},}\ }\href
  {https://doi.org/10.1016/0009-2614(93)89151-7} {\bibfield  {journal}
  {\bibinfo  {journal} {Chemical Physics Letters}\ }\textbf {\bibinfo {volume}
  {213}},\ \bibinfo {pages} {514–518} (\bibinfo {year} {1993})}\BibitemShut
  {NoStop}%
\bibitem [{\citenamefont {Rendell}\ and\ \citenamefont
  {Lee}(1994)}]{Rendell1994}%
  \BibitemOpen
  \bibfield  {author} {\bibinfo {author} {\bibfnamefont {A.~P.}\ \bibnamefont
  {Rendell}}\ and\ \bibinfo {author} {\bibfnamefont {T.~J.}\ \bibnamefont
  {Lee}},\ }\bibfield  {title} {\enquote {\bibinfo {title} {Coupled-cluster
  theory employing approximate integrals: An approach to avoid the input/output
  and storage bottlenecks},}\ }\href {https://doi.org/10.1063/1.468148}
  {\bibfield  {journal} {\bibinfo  {journal} {The Journal of Chemical Physics}\
  }\textbf {\bibinfo {volume} {101}},\ \bibinfo {pages} {400–408} (\bibinfo
  {year} {1994})}\BibitemShut {NoStop}%
\bibitem [{\citenamefont {Alml\"{o}f}(1991)}]{almlof1991}%
  \BibitemOpen
  \bibfield  {author} {\bibinfo {author} {\bibfnamefont {J.}~\bibnamefont
  {Alml\"{o}f}},\ }\bibfield  {title} {\enquote {\bibinfo {title} {Elimination
  of energy denominators in m{\o}ller—plesset perturbation theory by a
  laplace transform approach},}\ }\href
  {https://doi.org/10.1016/0009-2614(91)80078-C} {\bibfield  {journal}
  {\bibinfo  {journal} {Chemical Physics Letters}\ }\textbf {\bibinfo {volume}
  {181}},\ \bibinfo {pages} {319--320} (\bibinfo {year} {1991})}\BibitemShut
  {NoStop}%
\bibitem [{\citenamefont {H\"{a}ser}\ and\ \citenamefont
  {Alml\"{o}f}(1992)}]{Haser1992}%
  \BibitemOpen
  \bibfield  {author} {\bibinfo {author} {\bibfnamefont {M.}~\bibnamefont
  {H\"{a}ser}}\ and\ \bibinfo {author} {\bibfnamefont {J.}~\bibnamefont
  {Alml\"{o}f}},\ }\bibfield  {title} {\enquote {\bibinfo {title} {Laplace
  transform techniques in m{\o}ller-plesset perturbation theory},}\ }\href
  {https://doi.org/10.1063/1.462485} {\bibfield  {journal} {\bibinfo  {journal}
  {The Journal of Chemical Physics}\ }\textbf {\bibinfo {volume} {96}},\
  \bibinfo {pages} {489--494} (\bibinfo {year} {1992})}\BibitemShut {NoStop}%
\bibitem [{\citenamefont {Braess}\ and\ \citenamefont
  {Hackbusch}(2005)}]{braessApproximationExponentialSums2005}%
  \BibitemOpen
  \bibfield  {author} {\bibinfo {author} {\bibfnamefont {D.}~\bibnamefont
  {Braess}}\ and\ \bibinfo {author} {\bibfnamefont {W.}~\bibnamefont
  {Hackbusch}},\ }\bibfield  {title} {\enquote {\bibinfo {title} {Approximation
  of 1/x by exponential sums in [1, {$\infty$})},}\ }\href
  {https://doi.org/10.1093/imanum/dri015} {\bibfield  {journal} {\bibinfo
  {journal} {IMA J Numer Anal}\ }\textbf {\bibinfo {volume} {25}},\ \bibinfo
  {pages} {685--697} (\bibinfo {year} {2005})}\BibitemShut {NoStop}%
\bibitem [{\citenamefont {Constans}, \citenamefont {Ayala},\ and\ \citenamefont
  {Scuseria}(2000)}]{Constans2000}%
  \BibitemOpen
  \bibfield  {author} {\bibinfo {author} {\bibfnamefont {P.}~\bibnamefont
  {Constans}}, \bibinfo {author} {\bibfnamefont {P.~Y.}\ \bibnamefont
  {Ayala}},\ and\ \bibinfo {author} {\bibfnamefont {G.~E.}\ \bibnamefont
  {Scuseria}},\ }\bibfield  {title} {\enquote {\bibinfo {title} {Scaling
  reduction of the perturbative triples correction (t) to coupled cluster
  theory via laplace transform formalism},}\ }\href
  {https://doi.org/10.1063/1.1324989} {\bibfield  {journal} {\bibinfo
  {journal} {The Journal of Chemical Physics}\ }\textbf {\bibinfo {volume}
  {113}},\ \bibinfo {pages} {10451--10458} (\bibinfo {year}
  {2000})}\BibitemShut {NoStop}%
\bibitem [{\citenamefont {Kinoshita}, \citenamefont {Hino},\ and\ \citenamefont
  {Bartlett}(2003)}]{Kinoshita2003}%
  \BibitemOpen
  \bibfield  {author} {\bibinfo {author} {\bibfnamefont {T.}~\bibnamefont
  {Kinoshita}}, \bibinfo {author} {\bibfnamefont {O.}~\bibnamefont {Hino}},\
  and\ \bibinfo {author} {\bibfnamefont {R.~J.}\ \bibnamefont {Bartlett}},\
  }\bibfield  {title} {\enquote {\bibinfo {title} {Singular value decomposition
  approach for the approximate coupled-cluster method},}\ }\href
  {https://doi.org/10.1063/1.1609442} {\bibfield  {journal} {\bibinfo
  {journal} {The Journal of Chemical Physics}\ }\textbf {\bibinfo {volume}
  {119}},\ \bibinfo {pages} {7756--7762} (\bibinfo {year} {2003})}\BibitemShut
  {NoStop}%
\bibitem [{\citenamefont {Parrish}\ \emph {et~al.}(2019)\citenamefont
  {Parrish}, \citenamefont {Zhao}, \citenamefont {Hohenstein},\ and\
  \citenamefont {Mart{\'{\i}}nez}}]{Parrish2019}%
  \BibitemOpen
  \bibfield  {author} {\bibinfo {author} {\bibfnamefont {R.~M.}\ \bibnamefont
  {Parrish}}, \bibinfo {author} {\bibfnamefont {Y.}~\bibnamefont {Zhao}},
  \bibinfo {author} {\bibfnamefont {E.~G.}\ \bibnamefont {Hohenstein}},\ and\
  \bibinfo {author} {\bibfnamefont {T.~J.}\ \bibnamefont {Mart{\'{\i}}nez}},\
  }\bibfield  {title} {\enquote {\bibinfo {title} {Rank reduced coupled cluster
  theory. i. ground state energies and wavefunctions},}\ }\href
  {https://doi.org/10.1063/1.5092505} {\bibfield  {journal} {\bibinfo
  {journal} {The Journal of Chemical Physics}\ }\textbf {\bibinfo {volume}
  {150}},\ \bibinfo {pages} {164118} (\bibinfo {year} {2019})}\BibitemShut
  {NoStop}%
\bibitem [{\citenamefont {Hino}, \citenamefont {Kinoshita},\ and\ \citenamefont
  {Bartlett}(2004)}]{Hino2004}%
  \BibitemOpen
  \bibfield  {author} {\bibinfo {author} {\bibfnamefont {O.}~\bibnamefont
  {Hino}}, \bibinfo {author} {\bibfnamefont {T.}~\bibnamefont {Kinoshita}},\
  and\ \bibinfo {author} {\bibfnamefont {R.~J.}\ \bibnamefont {Bartlett}},\
  }\bibfield  {title} {\enquote {\bibinfo {title} {Singular value decomposition
  applied to the compression of t3 amplitude for the coupled cluster method},}\
  }\href {https://doi.org/10.1063/1.1763575} {\bibfield  {journal} {\bibinfo
  {journal} {The Journal of Chemical Physics}\ }\textbf {\bibinfo {volume}
  {121}},\ \bibinfo {pages} {1206--1213} (\bibinfo {year} {2004})}\BibitemShut
  {NoStop}%
\bibitem [{\citenamefont {Urban}\ \emph {et~al.}(1985)\citenamefont {Urban},
  \citenamefont {Noga}, \citenamefont {Cole},\ and\ \citenamefont
  {Bartlett}}]{Urban-JCP-1985}%
  \BibitemOpen
  \bibfield  {author} {\bibinfo {author} {\bibfnamefont {M.}~\bibnamefont
  {Urban}}, \bibinfo {author} {\bibfnamefont {J.}~\bibnamefont {Noga}},
  \bibinfo {author} {\bibfnamefont {S.~J.}\ \bibnamefont {Cole}},\ and\
  \bibinfo {author} {\bibfnamefont {R.~J.}\ \bibnamefont {Bartlett}},\
  }\bibfield  {title} {\enquote {\bibinfo {title} {{Towards a full CCSDT model
  for electron correlation}},}\ }\href {https://doi.org/10.1063/1.449067}
  {\bibfield  {journal} {\bibinfo  {journal} {The Journal of Chemical Physics}\
  }\textbf {\bibinfo {volume} {83}},\ \bibinfo {pages} {4041--4046} (\bibinfo
  {year} {1985})}\BibitemShut {NoStop}%
\bibitem [{\citenamefont {Lesiuk}(2019{\natexlab{a}})}]{Lesiuk2019}%
  \BibitemOpen
  \bibfield  {author} {\bibinfo {author} {\bibfnamefont {M.}~\bibnamefont
  {Lesiuk}},\ }\bibfield  {title} {\enquote {\bibinfo {title} {Efficient
  singular-value decomposition of the coupled-cluster triple excitation
  amplitudes},}\ }\href {https://doi.org/10.1002/jcc.25788} {\bibfield
  {journal} {\bibinfo  {journal} {Journal of Computational Chemistry}\ }\textbf
  {\bibinfo {volume} {40}},\ \bibinfo {pages} {1319--1332} (\bibinfo {year}
  {2019}{\natexlab{a}})}\BibitemShut {NoStop}%
\bibitem [{\citenamefont {Lesiuk}(2019{\natexlab{b}})}]{Lesiukacs2019}%
  \BibitemOpen
  \bibfield  {author} {\bibinfo {author} {\bibfnamefont {M.}~\bibnamefont
  {Lesiuk}},\ }\bibfield  {title} {\enquote {\bibinfo {title} {Implementation
  of the coupled-cluster method with single, double, and triple excitations
  using tensor decompositions},}\ }\href
  {https://doi.org/10.1021/acs.jctc.9b00985} {\bibfield  {journal} {\bibinfo
  {journal} {Journal of Chemical Theory and Computation}\ }\textbf {\bibinfo
  {volume} {16}},\ \bibinfo {pages} {453--467} (\bibinfo {year}
  {2019}{\natexlab{b}})}\BibitemShut {NoStop}%
\bibitem [{\citenamefont {Lesiuk}(2022)}]{Lesiuk2022}%
  \BibitemOpen
  \bibfield  {author} {\bibinfo {author} {\bibfnamefont {M.}~\bibnamefont
  {Lesiuk}},\ }\bibfield  {title} {\enquote {\bibinfo {title} {Quintic-scaling
  rank-reduced coupled cluster theory with single and double excitations},}\
  }\href {https://doi.org/10.1063/5.0071916} {\bibfield  {journal} {\bibinfo
  {journal} {The Journal of Chemical Physics}\ }\textbf {\bibinfo {volume}
  {156}},\ \bibinfo {pages} {064103} (\bibinfo {year} {2022})}\BibitemShut
  {NoStop}%
\bibitem [{\citenamefont {Hohenstein}\ \emph {et~al.}(2012)\citenamefont
  {Hohenstein}, \citenamefont {Parrish}, \citenamefont {Sherrill},\ and\
  \citenamefont {Martínez}}]{Hohenstein-JCP-2012}%
  \BibitemOpen
  \bibfield  {author} {\bibinfo {author} {\bibfnamefont {E.~G.}\ \bibnamefont
  {Hohenstein}}, \bibinfo {author} {\bibfnamefont {R.~M.}\ \bibnamefont
  {Parrish}}, \bibinfo {author} {\bibfnamefont {C.~D.}\ \bibnamefont
  {Sherrill}},\ and\ \bibinfo {author} {\bibfnamefont {T.~J.}\ \bibnamefont
  {Martínez}},\ }\bibfield  {title} {\enquote {\bibinfo {title}
  {{Communication: Tensor hypercontraction. III. Least-squares tensor
  hypercontraction for the determination of correlated wavefunctions}},}\
  }\href {https://doi.org/10.1063/1.4768241} {\bibfield  {journal} {\bibinfo
  {journal} {The Journal of Chemical Physics}\ }\textbf {\bibinfo {volume}
  {137}},\ \bibinfo {pages} {221101} (\bibinfo {year} {2012})}\BibitemShut
  {NoStop}%
\bibitem [{\citenamefont {Matthews}(2021)}]{Matthews-JCP-2021}%
  \BibitemOpen
  \bibfield  {author} {\bibinfo {author} {\bibfnamefont {D.~A.}\ \bibnamefont
  {Matthews}},\ }\bibfield  {title} {\enquote {\bibinfo {title} {{A critical
  analysis of least-squares tensor hypercontraction applied to MP3}},}\ }\href
  {https://doi.org/10.1063/5.0038764} {\bibfield  {journal} {\bibinfo
  {journal} {The Journal of Chemical Physics}\ }\textbf {\bibinfo {volume}
  {154}},\ \bibinfo {pages} {134102} (\bibinfo {year} {2021})}\BibitemShut
  {NoStop}%
\bibitem [{\citenamefont {Jiang}, \citenamefont {Turney},\ and\ \citenamefont
  {Schaefer}(2023{\natexlab{a}})}]{Jiang2023}%
  \BibitemOpen
  \bibfield  {author} {\bibinfo {author} {\bibfnamefont {A.}~\bibnamefont
  {Jiang}}, \bibinfo {author} {\bibfnamefont {J.~M.}\ \bibnamefont {Turney}},\
  and\ \bibinfo {author} {\bibfnamefont {H.~F.}\ \bibnamefont {Schaefer}},\
  }\bibfield  {title} {\enquote {\bibinfo {title} {Tensor hypercontraction form
  of the perturbative triples energy in coupled-cluster theory},}\ }\href
  {https://doi.org/10.1021/acs.jctc.2c00996} {\bibfield  {journal} {\bibinfo
  {journal} {Journal of Chemical Theory and Computation}\ }\textbf {\bibinfo
  {volume} {19}},\ \bibinfo {pages} {1476--1486} (\bibinfo {year}
  {2023}{\natexlab{a}})}\BibitemShut {NoStop}%
\bibitem [{\citenamefont {Pople}, \citenamefont {Head-Gordon},\ and\
  \citenamefont {Raghavachari}(1987)}]{Pople1987}%
  \BibitemOpen
  \bibfield  {author} {\bibinfo {author} {\bibfnamefont {J.~A.}\ \bibnamefont
  {Pople}}, \bibinfo {author} {\bibfnamefont {M.}~\bibnamefont {Head-Gordon}},\
  and\ \bibinfo {author} {\bibfnamefont {K.}~\bibnamefont {Raghavachari}},\
  }\bibfield  {title} {\enquote {\bibinfo {title} {Quadratic configuration
  interaction. a general technique for determining electron correlation
  energies},}\ }\href {https://doi.org/10.1063/1.453520} {\bibfield  {journal}
  {\bibinfo  {journal} {The Journal of Chemical Physics}\ }\textbf {\bibinfo
  {volume} {87}},\ \bibinfo {pages} {5968--5975} (\bibinfo {year}
  {1987})}\BibitemShut {NoStop}%
\bibitem [{\citenamefont {Kendall}\ and\ \citenamefont
  {Fr\"{u}chtl}(1997)}]{Kendall1997}%
  \BibitemOpen
  \bibfield  {author} {\bibinfo {author} {\bibfnamefont {R.~A.}\ \bibnamefont
  {Kendall}}\ and\ \bibinfo {author} {\bibfnamefont {H.~A.}\ \bibnamefont
  {Fr\"{u}chtl}},\ }\bibfield  {title} {\enquote {\bibinfo {title} {The impact
  of the resolution of the identity approximate integral method on modern ab
  initio algorithm development},}\ }\href
  {https://doi.org/10.1007/s002140050249} {\bibfield  {journal} {\bibinfo
  {journal} {Theoretical Chemistry Accounts: Theory, Computation, and Modeling
  (Theoretica Chimica Acta)}\ }\textbf {\bibinfo {volume} {97}},\ \bibinfo
  {pages} {158–163} (\bibinfo {year} {1997})}\BibitemShut {NoStop}%
\bibitem [{\citenamefont {Eichkorn}\ \emph {et~al.}(1995)\citenamefont
  {Eichkorn}, \citenamefont {Treutler}, \citenamefont {\"{O}hm}, \citenamefont
  {H\"{a}ser},\ and\ \citenamefont {Ahlrichs}}]{Eichkorn1995}%
  \BibitemOpen
  \bibfield  {author} {\bibinfo {author} {\bibfnamefont {K.}~\bibnamefont
  {Eichkorn}}, \bibinfo {author} {\bibfnamefont {O.}~\bibnamefont {Treutler}},
  \bibinfo {author} {\bibfnamefont {H.}~\bibnamefont {\"{O}hm}}, \bibinfo
  {author} {\bibfnamefont {M.}~\bibnamefont {H\"{a}ser}},\ and\ \bibinfo
  {author} {\bibfnamefont {R.}~\bibnamefont {Ahlrichs}},\ }\bibfield  {title}
  {\enquote {\bibinfo {title} {Auxiliary basis sets to approximate coulomb
  potentials (chem. phys. letters 240 (1995) 283-290)},}\ }\href
  {https://doi.org/10.1016/0009-2614(95)00838-u} {\bibfield  {journal}
  {\bibinfo  {journal} {Chemical Physics Letters}\ }\textbf {\bibinfo {volume}
  {242}},\ \bibinfo {pages} {652–660} (\bibinfo {year} {1995})}\BibitemShut
  {NoStop}%
\bibitem [{\citenamefont {Bernholdt}\ and\ \citenamefont
  {Harrison}(1998)}]{Bernholdt1998}%
  \BibitemOpen
  \bibfield  {author} {\bibinfo {author} {\bibfnamefont {D.~E.}\ \bibnamefont
  {Bernholdt}}\ and\ \bibinfo {author} {\bibfnamefont {R.~J.}\ \bibnamefont
  {Harrison}},\ }\bibfield  {title} {\enquote {\bibinfo {title} {Fitting basis
  sets for the ri-mp2 approximate second-order many-body perturbation theory
  method},}\ }\href {https://doi.org/10.1063/1.476732} {\bibfield  {journal}
  {\bibinfo  {journal} {The Journal of Chemical Physics}\ }\textbf {\bibinfo
  {volume} {109}},\ \bibinfo {pages} {1593–1600} (\bibinfo {year}
  {1998})}\BibitemShut {NoStop}%
\bibitem [{\citenamefont {Weigend}, \citenamefont {K\"{o}hn},\ and\
  \citenamefont {H\"{a}ttig}(2002)}]{Weigend2002_jcp}%
  \BibitemOpen
  \bibfield  {author} {\bibinfo {author} {\bibfnamefont {F.}~\bibnamefont
  {Weigend}}, \bibinfo {author} {\bibfnamefont {A.}~\bibnamefont {K\"{o}hn}},\
  and\ \bibinfo {author} {\bibfnamefont {C.}~\bibnamefont {H\"{a}ttig}},\
  }\bibfield  {title} {\enquote {\bibinfo {title} {Efficient use of the
  correlation consistent basis sets in resolution of the identity mp2
  calculations},}\ }\href {https://doi.org/10.1063/1.1445115} {\bibfield
  {journal} {\bibinfo  {journal} {The Journal of Chemical Physics}\ }\textbf
  {\bibinfo {volume} {116}},\ \bibinfo {pages} {3175–3183} (\bibinfo {year}
  {2002})}\BibitemShut {NoStop}%
\bibitem [{\citenamefont {Weigend}(2002)}]{Weigend2002}%
  \BibitemOpen
  \bibfield  {author} {\bibinfo {author} {\bibfnamefont {F.}~\bibnamefont
  {Weigend}},\ }\bibfield  {title} {\enquote {\bibinfo {title} {A fully direct
  ri-hf algorithm: Implementation, optimised auxiliary basis sets,
  demonstration of accuracy and efficiency},}\ }\href
  {https://doi.org/10.1039/b204199p} {\bibfield  {journal} {\bibinfo  {journal}
  {Physical Chemistry Chemical Physics}\ }\textbf {\bibinfo {volume} {4}},\
  \bibinfo {pages} {4285–4291} (\bibinfo {year} {2002})}\BibitemShut
  {NoStop}%
\bibitem [{\citenamefont {F.~Bell}\ and\ \citenamefont
  {Head-Gordon}(2010)}]{Bell-MP-2010}%
  \BibitemOpen
  \bibfield  {author} {\bibinfo {author} {\bibfnamefont {D.~L.}\ \bibnamefont
  {F.~Bell}}\ and\ \bibinfo {author} {\bibfnamefont {M.}~\bibnamefont
  {Head-Gordon}},\ }\bibfield  {title} {\enquote {\bibinfo {title} {Higher
  order singular value decomposition in quantum chemistry},}\ }\href@noop {}
  {\bibfield  {journal} {\bibinfo  {journal} {Molecular Physics}\ }\textbf
  {\bibinfo {volume} {108}},\ \bibinfo {pages} {2759--2773} (\bibinfo {year}
  {2010})}\BibitemShut {NoStop}%
\bibitem [{\citenamefont {Constans}\ and\ \citenamefont
  {Scuseria}(2003)}]{constans2003}%
  \BibitemOpen
  \bibfield  {author} {\bibinfo {author} {\bibfnamefont {P.}~\bibnamefont
  {Constans}}\ and\ \bibinfo {author} {\bibfnamefont {G.~E.}\ \bibnamefont
  {Scuseria}},\ }\bibfield  {title} {\enquote {\bibinfo {title} {The laplace
  transform perturbative triples correction ansatz},}\ }\href
  {https://doi.org/10.1135/cccc20030357} {\bibfield  {journal} {\bibinfo
  {journal} {Collection of Czechoslovak Chemical Communications}\ }\textbf
  {\bibinfo {volume} {68}},\ \bibinfo {pages} {357--373} (\bibinfo {year}
  {2003})}\BibitemShut {NoStop}%
\bibitem [{\citenamefont {Kats}, \citenamefont {Usvyat},\ and\ \citenamefont
  {Sch\"{u}tz}(2008)}]{kats2008}%
  \BibitemOpen
  \bibfield  {author} {\bibinfo {author} {\bibfnamefont {D.}~\bibnamefont
  {Kats}}, \bibinfo {author} {\bibfnamefont {D.}~\bibnamefont {Usvyat}},\ and\
  \bibinfo {author} {\bibfnamefont {M.}~\bibnamefont {Sch\"{u}tz}},\ }\bibfield
   {title} {\enquote {\bibinfo {title} {On the use of the laplace transform in
  local correlation methods},}\ }\href {https://doi.org/10.1039/B802993H}
  {\bibfield  {journal} {\bibinfo  {journal} {Phys. Chem. Chem. Phys.}\
  }\textbf {\bibinfo {volume} {10}},\ \bibinfo {pages} {3430--3439} (\bibinfo
  {year} {2008})}\BibitemShut {NoStop}%
\bibitem [{\citenamefont {Jiang}, \citenamefont {Turney},\ and\ \citenamefont
  {Schaefer}(2023{\natexlab{b}})}]{Jiang-JCTC-2023}%
  \BibitemOpen
  \bibfield  {author} {\bibinfo {author} {\bibfnamefont {A.}~\bibnamefont
  {Jiang}}, \bibinfo {author} {\bibfnamefont {J.~M.}\ \bibnamefont {Turney}},\
  and\ \bibinfo {author} {\bibfnamefont {H.~F.~I.}\ \bibnamefont {Schaefer}},\
  }\bibfield  {title} {\enquote {\bibinfo {title} {Tensor hypercontraction form
  of the perturbative triples energy in coupled-cluster theory},}\ }\href
  {https://doi.org/10.1021/acs.jctc.2c00996} {\bibfield  {journal} {\bibinfo
  {journal} {Journal of Chemical Theory and Computation}\ }\textbf {\bibinfo
  {volume} {19}},\ \bibinfo {pages} {1476--1486} (\bibinfo {year}
  {2023}{\natexlab{b}})},\ \bibinfo {note} {pMID: 36802552}\BibitemShut
  {NoStop}%
\bibitem [{\citenamefont {Karton}, \citenamefont {Sylvetsky},\ and\
  \citenamefont {Martin}(2017)}]{Karton2017}%
  \BibitemOpen
  \bibfield  {author} {\bibinfo {author} {\bibfnamefont {A.}~\bibnamefont
  {Karton}}, \bibinfo {author} {\bibfnamefont {N.}~\bibnamefont {Sylvetsky}},\
  and\ \bibinfo {author} {\bibfnamefont {J.~M.~L.}\ \bibnamefont {Martin}},\
  }\bibfield  {title} {\enquote {\bibinfo {title} {W4-17: A diverse and
  high-confidence dataset of atomization energies for benchmarking high-level
  electronic structure methods},}\ }\href {https://doi.org/10.1002/jcc.24854}
  {\bibfield  {journal} {\bibinfo  {journal} {Journal of Computational
  Chemistry}\ }\textbf {\bibinfo {volume} {38}},\ \bibinfo {pages} {2063--2075}
  (\bibinfo {year} {2017})}\BibitemShut {NoStop}%
\bibitem [{\citenamefont {Dunning}(1989)}]{dunningbasis}%
  \BibitemOpen
  \bibfield  {author} {\bibinfo {author} {\bibfnamefont {T.~H.}\ \bibnamefont
  {Dunning}},\ }\bibfield  {title} {\enquote {\bibinfo {title} {Gaussian basis
  sets for use in correlated molecular calculations. i. the atoms boron through
  neon and hydrogen},}\ }\href {https://doi.org/10.1063/1.456153} {\bibfield
  {journal} {\bibinfo  {journal} {The Journal of Chemical Physics}\ }\textbf
  {\bibinfo {volume} {90}},\ \bibinfo {pages} {1007--1023} (\bibinfo {year}
  {1989})}\BibitemShut {NoStop}%
\bibitem [{\citenamefont {Kendall}, \citenamefont {Dunning},\ and\
  \citenamefont {Harrison}(1992)}]{Kendall1992}%
  \BibitemOpen
  \bibfield  {author} {\bibinfo {author} {\bibfnamefont {R.~A.}\ \bibnamefont
  {Kendall}}, \bibinfo {author} {\bibfnamefont {T.~H.}\ \bibnamefont
  {Dunning}},\ and\ \bibinfo {author} {\bibfnamefont {R.~J.}\ \bibnamefont
  {Harrison}},\ }\bibfield  {title} {\enquote {\bibinfo {title} {Electron
  affinities of the first-row atoms revisited. systematic basis sets and wave
  functions},}\ }\href {https://doi.org/10.1063/1.462569} {\bibfield  {journal}
  {\bibinfo  {journal} {The Journal of Chemical Physics}\ }\textbf {\bibinfo
  {volume} {96}},\ \bibinfo {pages} {6796–6806} (\bibinfo {year}
  {1992})}\BibitemShut {NoStop}%
\bibitem [{\citenamefont {Woon}\ and\ \citenamefont
  {Dunning}(1995)}]{Woon1995}%
  \BibitemOpen
  \bibfield  {author} {\bibinfo {author} {\bibfnamefont {D.~E.}\ \bibnamefont
  {Woon}}\ and\ \bibinfo {author} {\bibfnamefont {T.~H.}\ \bibnamefont
  {Dunning}},\ }\bibfield  {title} {\enquote {\bibinfo {title} {Gaussian basis
  sets for use in correlated molecular calculations. v. core-valence basis sets
  for boron through neon},}\ }\href {https://doi.org/10.1063/1.470645}
  {\bibfield  {journal} {\bibinfo  {journal} {The Journal of Chemical Physics}\
  }\textbf {\bibinfo {volume} {103}},\ \bibinfo {pages} {4572–4585} (\bibinfo
  {year} {1995})}\BibitemShut {NoStop}%
\bibitem [{\citenamefont {Feyereisen}, \citenamefont {Fitzgerald},\ and\
  \citenamefont {Komornicki}(1993)}]{Feyereisen1993}%
  \BibitemOpen
  \bibfield  {author} {\bibinfo {author} {\bibfnamefont {M.}~\bibnamefont
  {Feyereisen}}, \bibinfo {author} {\bibfnamefont {G.}~\bibnamefont
  {Fitzgerald}},\ and\ \bibinfo {author} {\bibfnamefont {A.}~\bibnamefont
  {Komornicki}},\ }\bibfield  {title} {\enquote {\bibinfo {title} {Use of
  approximate integrals in ab initio theory. an application in mp2 energy
  calculations},}\ }\href {https://doi.org/10.1016/0009-2614(93)87156-w}
  {\bibfield  {journal} {\bibinfo  {journal} {Chemical Physics Letters}\
  }\textbf {\bibinfo {volume} {208}},\ \bibinfo {pages} {359–363} (\bibinfo
  {year} {1993})}\BibitemShut {NoStop}%
\bibitem [{\citenamefont {Dunlap}(2000)}]{dunlapdf}%
  \BibitemOpen
  \bibfield  {author} {\bibinfo {author} {\bibfnamefont {B.~I.}\ \bibnamefont
  {Dunlap}},\ }\bibfield  {title} {\enquote {\bibinfo {title} {Robust and
  variational fitting},}\ }\href {https://doi.org/10.1039/B000027M} {\bibfield
  {journal} {\bibinfo  {journal} {Phys. Chem. Chem. Phys.}\ }\textbf {\bibinfo
  {volume} {2}},\ \bibinfo {pages} {2113--2116} (\bibinfo {year}
  {2000})}\BibitemShut {NoStop}%
\bibitem [{\citenamefont {Jung}\ \emph {et~al.}(2005)\citenamefont {Jung},
  \citenamefont {Sodt}, \citenamefont {Gill},\ and\ \citenamefont
  {Head-Gordon}}]{auxbasis}%
  \BibitemOpen
  \bibfield  {author} {\bibinfo {author} {\bibfnamefont {Y.}~\bibnamefont
  {Jung}}, \bibinfo {author} {\bibfnamefont {A.}~\bibnamefont {Sodt}}, \bibinfo
  {author} {\bibfnamefont {P.~M.~W.}\ \bibnamefont {Gill}},\ and\ \bibinfo
  {author} {\bibfnamefont {M.}~\bibnamefont {Head-Gordon}},\ }\bibfield
  {title} {\enquote {\bibinfo {title} {Auxiliary basis expansions for
  large-scale electronic structure calculations},}\ }\href
  {https://doi.org/10.1073/pnas.0408475102} {\bibfield  {journal} {\bibinfo
  {journal} {Proc. Natl. Acad. Sci.}\ }\textbf {\bibinfo {volume} {102}},\
  \bibinfo {pages} {6692--6697} (\bibinfo {year} {2005})}\BibitemShut {NoStop}%
\bibitem [{\citenamefont {Martin}(1996)}]{Martin1996}%
  \BibitemOpen
  \bibfield  {author} {\bibinfo {author} {\bibfnamefont {J.~M.}\ \bibnamefont
  {Martin}},\ }\bibfield  {title} {\enquote {\bibinfo {title} {Ab initio total
  atomization energies of small molecules {\textemdash} towards the basis set
  limit},}\ }\href {https://doi.org/10.1016/0009-2614(96)00898-6} {\bibfield
  {journal} {\bibinfo  {journal} {Chemical Physics Letters}\ }\textbf {\bibinfo
  {volume} {259}},\ \bibinfo {pages} {669--678} (\bibinfo {year}
  {1996})}\BibitemShut {NoStop}%
\bibitem [{\citenamefont {Matthews}\ \emph {et~al.}(2020)\citenamefont
  {Matthews}, \citenamefont {Cheng}, \citenamefont {Harding}, \citenamefont
  {Lipparini}, \citenamefont {Stopkowicz}, \citenamefont {Jagau}, \citenamefont
  {Szalay}, \citenamefont {Gauss},\ and\ \citenamefont {Stanton}}]{cfour}%
  \BibitemOpen
  \bibfield  {author} {\bibinfo {author} {\bibfnamefont {D.~A.}\ \bibnamefont
  {Matthews}}, \bibinfo {author} {\bibfnamefont {L.}~\bibnamefont {Cheng}},
  \bibinfo {author} {\bibfnamefont {M.~E.}\ \bibnamefont {Harding}}, \bibinfo
  {author} {\bibfnamefont {F.}~\bibnamefont {Lipparini}}, \bibinfo {author}
  {\bibfnamefont {S.}~\bibnamefont {Stopkowicz}}, \bibinfo {author}
  {\bibfnamefont {T.-C.}\ \bibnamefont {Jagau}}, \bibinfo {author}
  {\bibfnamefont {P.~G.}\ \bibnamefont {Szalay}}, \bibinfo {author}
  {\bibfnamefont {J.}~\bibnamefont {Gauss}},\ and\ \bibinfo {author}
  {\bibfnamefont {J.~F.}\ \bibnamefont {Stanton}},\ }\bibfield  {title}
  {\enquote {\bibinfo {title} {Coupled-cluster techniques for computational
  chemistry: The cfour program package},}\ }\href
  {https://doi.org/10.1063/5.0004837} {\bibfield  {journal} {\bibinfo
  {journal} {The Journal of Chemical Physics}\ }\textbf {\bibinfo {volume}
  {152}},\ \bibinfo {pages} {214108} (\bibinfo {year} {2020})}\BibitemShut
  {NoStop}%
\end{thebibliography}%

\begin{figure}[]
    \centering
    \includegraphics[width=0.5\textwidth]{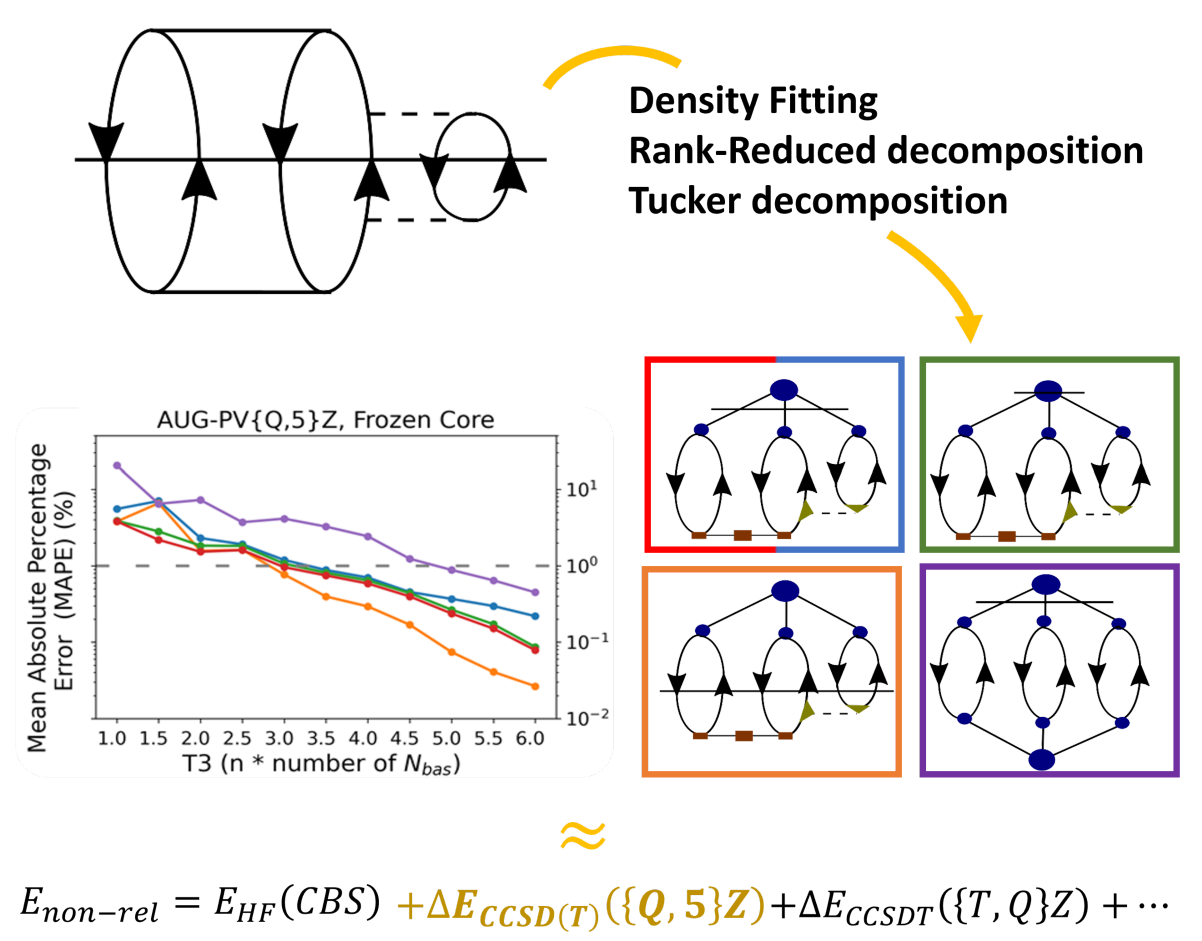}
    \caption{Table of contents figure.}
\end{figure}

\end{document}